\documentclass[twocolumn,showpacs]{revtex4}
\usepackage{graphicx}
\usepackage{dcolumn}

\begin{document}

%\preprint{APS/123-QED}

\title{Particle alignments and shape change
       in $^{66}$Ge and $^{68}$Ge}

\author{M. Hasegawa$^{1}$, K. Kaneko$^{2}$, and T. Mizusaki$^{3}$}

\affiliation{
$^{1}$Laboratory of Physics, Fukuoka Dental College, Fukuoka 814-0193, Japan \\
$^{2}$Department of Physics, Kyushu Sangyo University, Fukuoka 813-8503, Japan \\
$^{3}$Institute of Natural Sciences, Senshu University, Kawasaki, Kanagawa,
 214-8580, Japan
}

\date{\today}

\begin{abstract}

  The structure of the $N \approx Z$ nuclei $^{66}$Ge and $^{68}$Ge
 is studied by the shell model on a spherical basis. The calculations
 with an extended $P+QQ$ Hamiltonian in the configuration space
 ($2p_{3/2}$, $1f_{5/2}$, $2p_{1/2}$, $1g_{9/2}$) succeed in reproducing
 experimental energy levels, moments of inertia and $Q$ moments
 in Ge isotopes.  Using the reliable wave functions, this paper investigates
 particle alignments and nuclear shapes in $^{66}$Ge and $^{68}$Ge.
 It is shown that structural changes in the four sequences of the positive-
 and negative-parity yrast states with even $J$ and odd $J$ are caused
 by various types of particle alignments in the $g_{9/2}$ orbit.
 The nuclear shape is investigated by calculating spectroscopic $Q$
 moments of the first and second $2^+$ states, and moreover
 the triaxiality is examined by the constrained Hatree-Fock method. 
 The changes of the first band crossing and the nuclear deformation
 depending on the neutron number are discussed.

\end{abstract}


\maketitle

%=================================================================
\section{Introduction}\label{sec1}

   The Ge isotopes $^{66}$Ge and $^{68}$Ge have been studied intensively
in experiments (\cite{Nolte,Cleenmann,Sound,Boucenna,Hermkens,Stefanova}
for $^{66}$Ge and \cite{Pardo,Lima,Chat1,Chat2,Ward} for $^{68}$Ge)
and in theoretical investigations
\cite{Stefanova,Chat1,Ward,Barclay,Petrovici1,Petrovici2,Petrovici3,
Hsieh,Chuu,Elliott} for many years.
A thorough theoretical investigation was carried out by Petrovici et al.
with the excited VAMPIR (variation after mean field projection
in realistic model spaces) method for $^{68}$Ge
 \cite{Chat1,Petrovici1,Petrovici2,Petrovici3}.
The recent development of experimental techniques has accomplished
detailed measurements of $^{66}$Ge \cite{Stefanova} and $^{68}$Ge
\cite{Ward}.  The experiments have found several bands with positive and
negative parities up to high spins ($J \le 28$).  A hot topic of
the Ge isotopes has been the coexistence of oblate and prolate shapes
\cite{Bengtsson,Nazarewicz,Sarri} and possible $\gamma$ softness
 \cite{Ennis,Yamagami}.  The calculations based on the deformed mean
field approximation in Refs. \cite{Stefanova,Sarri} predict oblate shapes
for low-lying states of the ground-state bands of $^{66}$Ge and $^{68}$Ge,
which corresponds with the prediction of the excited VAMPIR calculations
for $^{68}$Ge.  Other theoretical approaches
 \cite{Barclay,Hsieh,Chuu,Elliott}, however, do not necessarily provide
the same explanation as the mean field picture.
The detailed data \cite{Stefanova,Ward} demand to make further theoretical
investigations into the structure of not only the low-energy states
but the high-spin states.

   We carried out shell model calculations on a spherical basis
for the $^{64}$Ge nucleus in a previous paper \cite{Kaneko}.
The calculations with the extended $P+QQ$ Hamiltonian \cite{Hase1,Hase2}
in the configuration space ($2p_{3/2}$, $1f_{5/2}$, $2p_{1/2}$, $1g_{9/2}$)
successfully described characteristics of the structure of $^{64}$Ge. 
The shell model has advantages that the nuclear deformation is dynamically
determined through nuclear interactions and wave functions are strictly
determined, which makes it possible to calculate various physical quantities
and to discuss the structure of bands in detail.  The shell model is
expected to be fruitful for the investigation into the recent data
on $^{66}$Ge and $^{68}$Ge. 

   For $^{64}$Ge, $^{65}$Ge, $^{66}$Ge, $^{67}$Ge, $^{68}$Ge, and $^{70}$Ge,
we have carried out large-scale shell model calculations using the calculation
code \cite{Mizusaki}.  The calculations reproduce well experimentally
observed energy levels and other properties of $^{66}$Ge and $^{68}$Ge.
In a recent paper \cite{Hase4}, we have reported an interesting feature
on the structural change of the even-$J$ positive-parity yrast states,
i.e., successive three bands with different types of particle alignments
(including proton-neutron alignment) in the $g_{9/2}$ orbit.
We show, in this paper, that similar changes are caused by various types
of particle alignments in the $g_{9/2}$ orbit, also in the other sequences of
positive-parity yrast states with odd $J$ and negative-parity yrast states
with even $J$ and odd $J$.
Spectroscopic $Q$ moments calculated with the reliable wave functions
provide useful information about the nuclear shape and shape change
depending on the neutron number.  In order to investigate the shape 
and $\gamma$ softness further, we calculate the potential energy surface
in the plane of the deformation parameter $q$ and the angle $\gamma$
using the constrained Hartree-Fock (CHF) method too.

   Section~\ref{sec2} presents parameters of the extended $P+QQ$ Hamiltonian
determined for $^{65,66,67,68}$Ge and energy levels obtained.
In section~\ref{sec3}, experimental graphs of spin versus angular frequency
and electromagnetic transition probabilities $B(E2)$ and $B(M1)$ are compared
with the shell model results.
In section~\ref{sec4}, various types of particle alignments are investigated
in the four sequences of positive- and negative-parity yrast states
with even $J$ and odd $J$.  We discuss the change of the first band 
crossing in the Ge and Zn isotopes in section~\ref{sec5}.
Section~\ref{sec6} investigates the nuclear shape and the shape change.
A summary is given in section~\ref{sec7}.

%=================================================================
\section{Level schemes}\label{sec2}

%===============  fig. 1  ========================================
\begin{figure}[b]
\includegraphics[width=8.0cm,height=10.2cm]{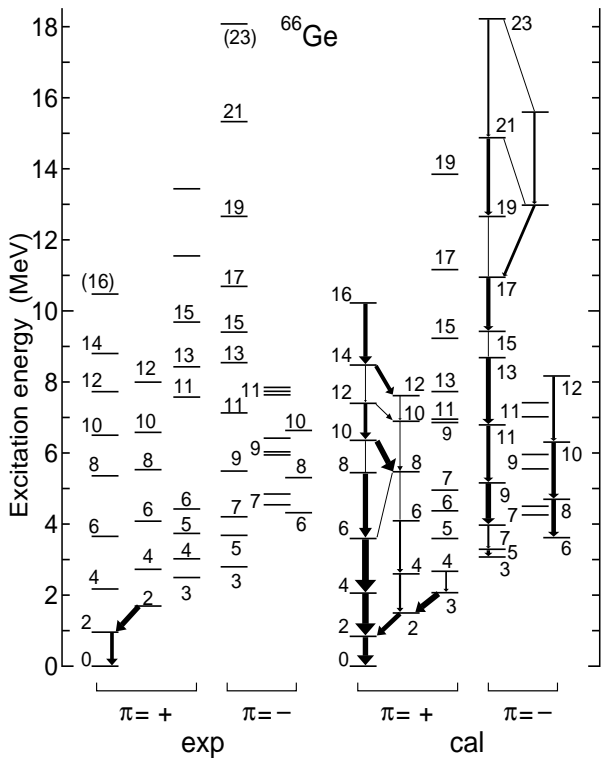}
  \caption{Experimental and calculated energy levels of $^{66}$Ge.
           The widths of the arrows denote relative values of $B(E2)$.}
  \label{fig1}
\end{figure}
%=================================================================

  The extended $P+QQ$ Hamiltonian is composed of the single-particle energies,
monopole corrections, $J=0$ and $J=2$ isovector pairing forces,
quadrupole-quadrupole ($QQ$) force and octupole-octupole ($OO$) force
(see Refs. \cite{Hase1,Hase2} in detail):
\begin{eqnarray}
 H & = & H_{\rm sp} + H_{\rm mc}
        + H_{P_0} + H_{P_2} + H_{QQ} + H_{OO}  \nonumber \\
   & = & \sum_{\alpha} \varepsilon_a c_\alpha^\dag c_\alpha + H_{\rm mc}
       -\sum_{J=0,2} \frac{1}{2} g_J \sum_{M\kappa} P^\dag_{JM1\kappa} P_{JM1\kappa}
          \nonumber \\
   & - & \frac{1}{2} \frac{\chi_2}{b^4} \sum_M :Q^\dag_{2M} Q_{2M}: 
         - \frac{1}{2} \frac{\chi_3}{b^6} \sum_M :O^\dag_{3M} O_{3M}:,
          \label{eq:1}
\end{eqnarray}
where the force strengths $\chi_2$ and $\chi_3$ are defined so as to have
the dimension of energy by excluding the harmonic-oscillator range parameter
$b$.  In Ref. \cite{Kaneko} for $^{64}$Ge, we employed
the same single-particle energies as those in Ref. \cite{Rudolph}
which are extracted from the energy levels of $^{57}$Ni
($\varepsilon_{g9/2} - \varepsilon_{p3/2} = 3.7$ MeV).
However, the parameters cannot reproduce the relative energies of
the positive- and negative-parity states in odd-mass Ge isotopes.
We therefore lowered the $g_{9/2}$ orbit toward the $pf$ shell so that
our shell model can reproduce experimental level schemes of odd-mass and
even-mass Ge isotopes, and also $^{66}$As as a whole.
This was linked with the search for force strengths.
We thus obtained the following set of parameters for the Ge isotopes.
The single-particle energies are
\begin{eqnarray}
 & {} & \varepsilon_{p3/2} = 0.00,  \quad \varepsilon_{f5/2} = 0.77,  \nonumber \\
 & {} & \varepsilon_{p1/2} = 1.11, \quad \varepsilon_{g9/2} = 2.50
  \quad \mbox{ in MeV}.  \label{eq:2}
\end{eqnarray}
The value $\varepsilon_{g9/2}=2.5$ MeV coincides with that of Ref. \cite{Aberg}.
The single-particle energies are discussed in Ref. \cite{Sun2}.
The force strengths determined for $^{66}$Ge are
\begin{eqnarray}
 & {} &  g_0 = 0.27(64/A), \quad g_2 = 0.0,            \nonumber \\
 & {} &  \chi_2 = 0.25(64/A)^{5/3},
         \chi_3 = 0.05(64/A)^2 \mbox{ in MeV}, \label{eq:3}
\end{eqnarray}
with $A=66$. We can get good results for $^{68}$Ge using the force strengths
(\ref{eq:3}) with $A=68$.  However, a little better results are obtained
by setting $A=66$ in Eq. (\ref{eq:3}) also for $^{68}$Ge.
We therefore use the fixed force parameters (values of Eq. (\ref{eq:3})
with $A=66$) for the Ge isotopes in this paper.
 The adopted monopole corrections are:
\begin{eqnarray}
 & {} & H^{T=1}_{\rm mc}(p_{3/2},f_{5/2}) = -0.3,
   \quad H^{T=1}_{\rm mc}(p_{3/2},p_{1/2}) = -0.3, \nonumber \\
 & {} & H^{T=1}_{\rm mc}(f_{5/2},p_{1/2}) = -0.4,
   \quad H^{T=1}_{\rm mc}(g_{9/2},g_{9/2}) = -0.2, \nonumber \\
 & {} & H^{T=0}_{\rm mc}(g_{9/2},g_{9/2}) = -0.1   \quad \mbox{ in MeV}.
               \label{eq:4}
\end{eqnarray}

%===============  fig. 2  ========================================
\begin{figure}[b]
\includegraphics[width=8.4cm,height=12cm]{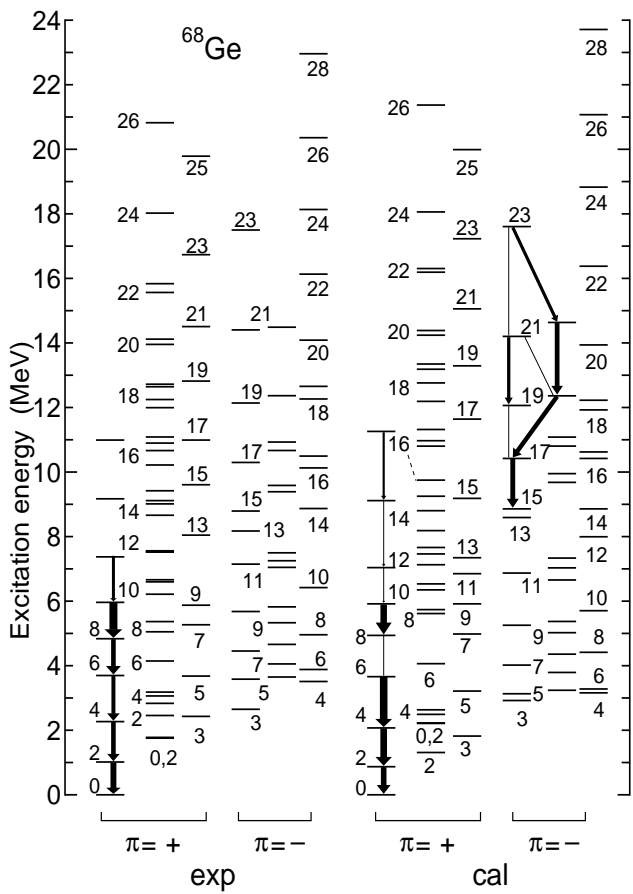}
  \caption{Experimental and calculated energy levels of $^{68}$Ge.
           The widths of the arrows denote relative values of $B(E2)$.}
  \label{fig2}
\end{figure}
%=================================================================

   Let us show energy levels of $^{66}$Ge and $^{68}$Ge calculated
with the parameters (\ref{eq:2}), (\ref{eq:3}), and (\ref{eq:4}), 
in Figs. \ref{fig1} and \ref{fig2}.  In these figures, the calculated results
excellently explain the energy levels observed in $^{66}$Ge and $^{68}$Ge
\cite{Stefanova,Ward}, reproducing several sequences of the positive- and
negative-parity states (except for a superdeformed band of $^{68}$Ge
which is not shown in Fig. \ref{fig2}). 
For $^{68}$Ge, detailed band scheme was proposed from $\Delta J=2$
electromagnetic transitions in the experiment \cite{Ward},
where, for instance, the $14_4^+$ and $16_6^+$ states are connected to
the yrast states $8_1^+$, $10_1^+$, and $12_1^+$.
This sequence of the states is shown in the most left columns of
the experimental and calculated results respectively in Fig. \ref{fig2}.
The calculated $14_3^+$ and $16_4^+$ states which are connected by large
$B(E2)$ values to the yrast $12_1^+$ state correspond well
to the experimental $14_4^+$ and $16_6^+$ states as shown in Fig. \ref{fig2}.
(Note that there are many levels with $J^\pi =14^+$ ($J^\pi =16^+$)
near $14_3^+$ ($16_4^+$) in the calculation.)
   We calculated also energy levels of $^{64}$Ge and $^{70}$Ge,
although high-spin states with $J^\pi \ge 10^+$ have not been detected.
Our model approximately reproduces the experimental energy levels
up to $8^+$ in $^{64}$Ge and $^{70}$Ge, as shown in Fig. \ref{fig14} later.

%===============  fig. 3  ========================================
\begin{figure}[b]
\includegraphics[width=8.2cm,height=9.8cm]{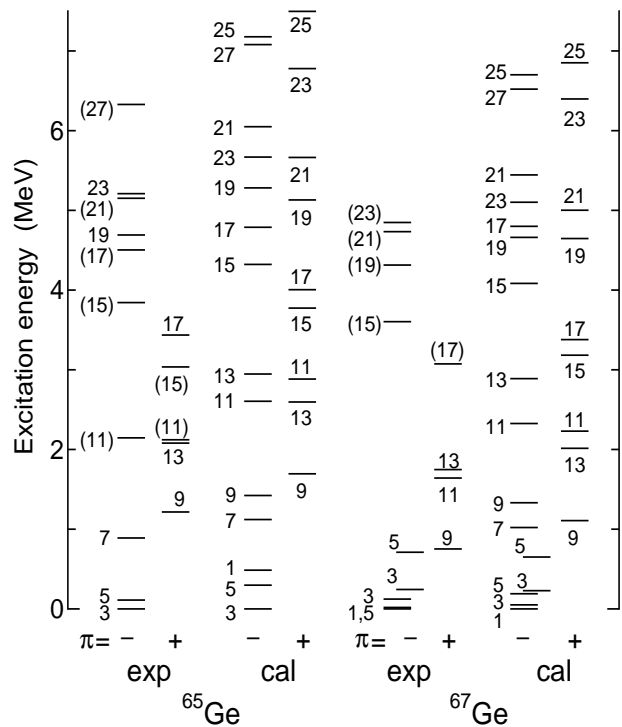}
  \caption{Experimental and calculated energy levels of $^{65}$Ge
           and $^{67}$Ge.
           The spin of each state is denoted by the double number $2J$.}
  \label{fig3}
\end{figure}
%=================================================================

  The extended $P+QQ$ model with the above parameters describes quite well
experimental energy levels of the odd-mass isotopes $^{65}$Ge and $^{67}$Ge
\cite{ENSDF}, as shown in Fig. \ref{fig3}.
 The energy levels of the positive- and negative-parity states
are approximately reproduced, although the agreement with experimental
ones in the odd-mass Ge isotopes is worse than that in the even-mass Ge
isotopes.  There are contradictions between theory and experiment
in the order of some energy levels with serial spins, and the calculated
high-spin states are pushed up as compared with the experimental ones.
However, our model correctly predicts the spins of
the ground states, $3/2^-$ for $^{65}$Ge and $1/2^-$ for $^{67}$Ge.
We can see one-to-one correspondence between the experimental and calculated
energy levels of $^{65}$Ge and $^{67}$Ge in Fig. \ref{fig3}.
Such a consistent description of both the even- and odd-mass Ge isotopes 
has not been reported.

   From these results, the extended $P+QQ$ interaction can be regarded
as an almost realistic effective interaction for the Ge isotopes.
The success in reproducing the level schemes testifies to the reliability
of the wave functions obtained.  We shall calculate various quantities
using the obtained wave functions and discuss the structure of $^{66}$Ge
and $^{68}$Ge in the following sections.

%=================================================================
\section{Moments of inertia and electromagnetic transitions}\label{sec3}

\subsection{$J - \omega$ graphs}\label{sec3.1}

   Illustrating the relation between the spin $J$ and the rotational
frequency $\omega (J)=(E(J)-E(J-2))/2$ with a graph (which we call 
the ``$J-\omega$ graph") is useful in seeing the change of nuclear structure,
because the moment of inertia $J/\omega (J)$ reflects competition
among various nuclear correlations.

\begin{figure}[t]
\includegraphics[width=7cm,height=12.0cm]{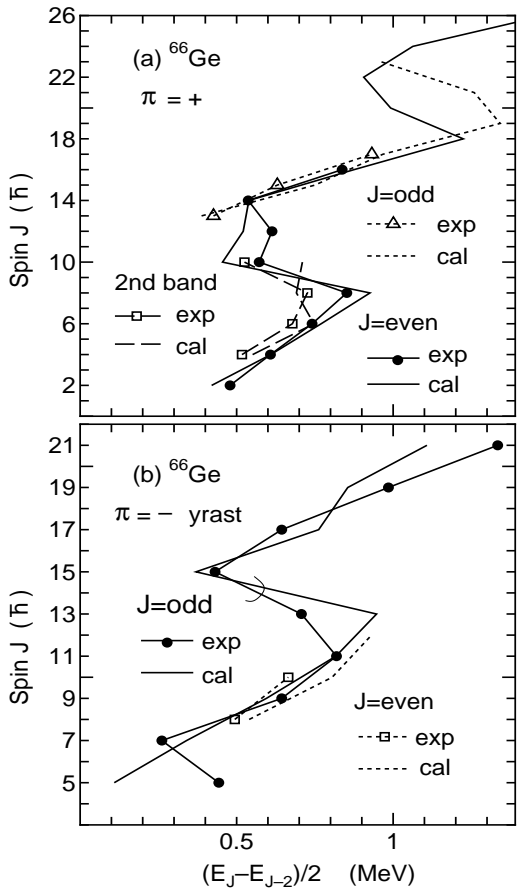}
  \caption{The $J-\omega$ graphs for $^{66}$Ge:
          (a) the positive-parity yrast states with even $J$ and odd $J$,
          and the second band on $2_2^+$;
          (b) the negative-parity yrast states with even $J$ and odd $J$.}
  \label{fig4}
\end{figure}

   Figure \ref{fig4}(a) shows the $J-\omega$ graphs for the three sequences
of positive-parity states of $^{66}$Ge which are illustrated in Fig. \ref{fig1}:
the even-$J$ yrast states including the ground-state ($gs$) band
up to $8_1^+$; the second band on the $2_2^+$ state;
the odd-$J$ yrast states on the $11_1^+$ state.
 Our model reproduces well the changes of the moments of inertia
for these three sequences of positive-parity states.
The agreement with the experimental moments of inertia is much better
than that of the total Routhian surface (TRS) calculations \cite{Stefanova}
and that of the projected shell model \cite{Sun2}.
This indicates that our wave functions are better than those of
the TRS calculations.

%===============  fig. 4  ========================================
%=================================================================

   In Fig. \ref{fig4}(a), the $J-\omega$ graph for the even-$J$ positive-parity
yrast states displays a stable (collective) rotation up to $8_1^+$ and
a sharp backbending toward $10_1^+$.  The remarkable backbending
from $8_1^+$ to $10_1^+$  indicates a structural change.
  We shall focus our attention on this phenomenon later.
The straight line starting from $14_1^+$ after the transitional
state $12_1^+$ is notable in the $J-\omega$ graph for the even-$J$ yrast
states in Fig. \ref{fig4}(a).
It is also interesting that the $J-\omega$ graph for the odd-$J$ states 
$13_1^+$, $15_1^+$, $17_1^+$, and $19_1^+$ is almost equal to the $J-\omega$
graph for the even-$J$ states $14_1^+$, $16_1^+$, and $18_1^+$.
The similar straight lines suggest that these states are generated
by the same rotor and the structure varies gradually.
The $J-\omega$ graphs in Fig. \ref{fig4}(a) predict structural changes
at $20_1^+$ in the sequence of the even-$J$ positive-parity yrast states
and at $21_1^+$ in the sequence of the odd-$J$ positive-parity yrast states.

   Figure \ref{fig4}(b) shows the $J-\omega$ graphs for the negative-parity
yrast states with odd $J$ and even $J$ of $^{66}$Ge.  Our model reproduces
well the change of the experimental moments of inertia.
 For the odd-$J$ yrast states, although there are deviations
from the experimental graph at $5_1^-$ and $13_1^-$, the sharp backbending
toward $15_1^-$ is well explained. 
 This backbending indicates a structural change at the $15_1^-$ state
in the sequence of the odd-$J$ negative-parity yrast states.
The $J-\omega$ graph for the odd-$J$ yrast states has a straight line
for the $15_1^-$, $17_1^-$, $19_1^-$, and $21_1^-$ states,
which is nearly equal to the two lines for the even-$J$ positive-parity
yrast states ($14_1^+$, $16_1^+$, $18_1^+$) and for the odd-$J$ ones
($13_1^+$, $15_1^+$, $17_1^+$, $19_1^+$) in Fig. \ref{fig4}(a).
These three bands ($15_1^- - 21_1^-$, $14_1^+ - 18_1^+$, and $13_1^+ - 19_1^+$)
seem to have a similar structure.
 Figure \ref{fig4}(b) also shows that the even-$J$ states $6_1^-$, 
$8_1^-$, and $10_1^-$ have a similar structure to the odd-$J$ states $7_1^-$, 
$9_1^-$, and $11_1^-$.

%===============  fig. 5  ========================================
\begin{figure}[b]
\includegraphics[width=7cm,height=12.0cm]{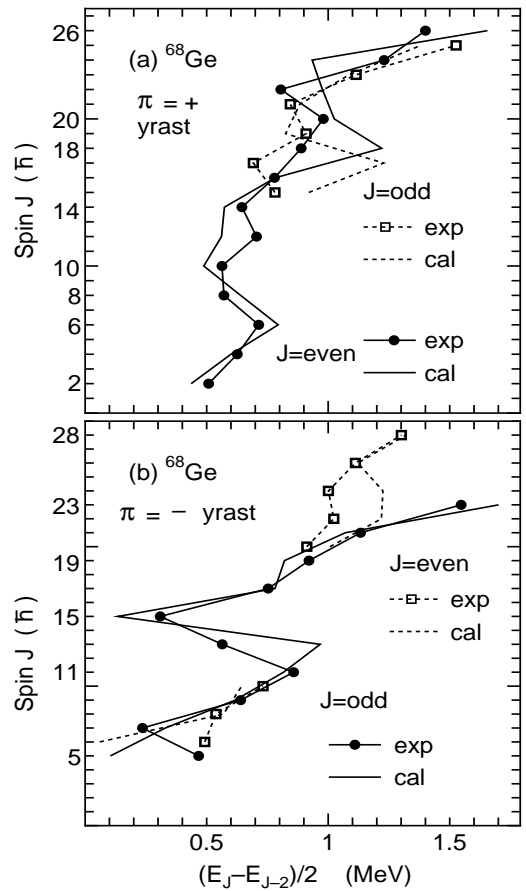}
  \caption{The $J-\omega$ graphs for $^{68}$Ge:
          (a) the positive-parity yrast states with even $J$ and odd $J$;
          (b) the negative-parity yrast states with even $J$ and odd $J$.}
  \label{fig5}
\end{figure}
%=================================================================

  We show the $J-\omega$ graphs for the positive-parity yrast states
with even $J$ and odd $J$ of $^{68}$Ge, in Fig. \ref{fig5}(a).  
Here we illustrate the $J-\omega$ graphs for the yrast states
instead of the sequence connected by $\Delta J=2$ transitions
(which is  shown in the most left columns of the experimental
and calculated energy levels in Fig. \ref{fig2}),
because we are concerned with how the structure changes
as the spin $J$ increases and wish to compare the $^{68}$Ge graphs
with the $^{66}$Ge graphs.

 In Fig. \ref{fig5}(a), our model reproduces quite well the experimental
$J-\omega$ graphs for $^{68}$Ge, except for $18_1^+$ and $17_1^+$. 
The comparison between Figs. \ref{fig4}(a) and \ref{fig5}(a) indicates
a similarity between the low-spin states up to $6_1^+$ of $^{66}$Ge
and $^{68}$Ge in the moments of inertia (hence the structure). 
There is, however, a significant difference between $^{66}$Ge and $^{68}$Ge.
The backbending takes place at $8_1^+$ in $^{68}$Ge, while it happens
at $10_1^+$ in $^{66}$Ge.  This difference is another interest discussed later.
The theoretical $J-\omega$ graph for the states $14_1^+$, $16_1^+$, and $18_1^+$
of $^{68}$Ge is similar to the graph for the corresponding states of $^{66}$Ge.
The experimental $J-\omega$ graph is different from the theoretical
one at $18_1^+$ in $^{68}$Ge.
The experiment \cite{Ward} observed a few sets of $14^+$, $16^+$, $18^+$, and
$20^+$ belonging to different bands.  The above discrepancy in the energy
spacing $E(18_1^+)-E(16_1^+)$ in Fig. \ref{fig5}(a) suggests a considerable
interplay between these bands near $16^+$ and $18^+$
which is not sufficiently taken into account in our model.
The two $J - \omega$ graphs for the experimental even-$J$ yrast states
on $18_1^+$ and odd-$J$ yrast states on $17_1^+$, which suggest a resemblance
between them, are well reproduced by the theory.

   The $J - \omega$ graphs for the negative-parity yrast states
with odd $J$ and even $J$ of $^{68}$Ge are illustrated in Fig. \ref{fig5}(b).
Our model is successful in describing the change of the moments of inertia
for the odd-$J$ yrast states, except for $5_1^-$ and $13_1^-$.
The sharp backbending toward $15_1^-$ is finely reproduced.
The low-spin band from $5_1^-$ to $11_1^-$ (except for the bad energy spacing
between $3_1^-$ and $5_1^-$) and the high-spin sequence on $15_1^-$ are also
reproduced well.  It is notable that the $J - \omega$ graph for the odd-$J$
negative-parity yrast states of $^{68}$Ge resembles that of $^{66}$Ge,
which suggests a similar structure of these states in $^{68}$Ge and
$^{66}$Ge.
In $^{68}$Ge, high-spin negative-parity yrast states with even $J$ are
experimentally observed  up to $28_1^-$ \cite{Ward}.
The moments of inertia of these states are well reproduced by the theory,
in Fig. \ref{fig5}(b).

\subsection{$B(E2)$ and $B(M1)$}\label{sec3.2}

   To investigate the correspondence between experimental and calculated bands,
we calculated reduced $E2$ transition probabilities using the effective charge
$e_p=1.5e$ for proton and $e_n=0.5e$ for neutron, and the harmonic-oscillator
range parameter $b=A^{1/3}$.  Experimental $B(E2)$ values and some of the
calculated $B(E2)$ values are denoted by the widths of the arrows
in Figs. \ref{fig1} and \ref{fig2}, and are tabulated in Table \ref{table1},
where the experimental data are taken from the evaluated nuclear structure data
file \cite{ENSDF}.

%===============  table 1  ========================
\begin{table}[b]
\caption{$B(E2)$ values for the positive-parity yrast states and 
         some collective states of $^{66}$Ge and $^{68}$Ge.
         Non-yrast states are distinguished with their subscripts
         from the yrast states with no subscript.
         The subscript denotes a serial number for each spin $J$.
         The most right column shows the $B(E2)$ values of
         Ref. \cite{Petrovici3}.}
\begin{tabular}{l|cc|ccc}   \hline
        & \multicolumn{2}{|c}{$^{66}$Ge  [$e^2$fm$^{4}$]}
         & \multicolumn{3}{|c}{$^{68}$Ge  [$e^2$fm$^{4}$]} \\ \hline
$J_i^+ \rightarrow J_f^+$ & exp & cal & exp & cal & \cite{Petrovici3} \\ \hline
$2 \rightarrow 0$       & $190 \pm 36$ & 281 & $292 \pm 33$ & 278  & 100 \\
$4 \rightarrow 2$       & $>$152 & 337 & $229 \pm 30$ & 330  & 563 \\
$6 \rightarrow 4$       & $>$19  & 351 & $198 \pm 66$ & 376  & 792 \\
$8 \rightarrow 6$       &        & 275 & $231 \pm 49$  & 0.09  & 291 \\
$8_2 \rightarrow 6$     &        &   4 & $198^{+66}_{-198}$ & 321  &     \\
$8 \rightarrow 6_2$     &        &   4 & $157 \pm 33$ & 0.02 &     \\
$8_2 \rightarrow 6_2$   &        & 0.05 & $297^{+99}_{-297}$ & 3.8  &     \\
$10 \rightarrow 8$      & $<$74  &   5 & $396^{+82}_{-396}$ & 322  & 852 \\
$10_2 \rightarrow 8$    &        & 1.8 & $>$25  & 2.7  &     \\
$10 \rightarrow 8_2$    & $<$206 & 285 &        & 0.01 &     \\
$10_2 \rightarrow 8_2$  &        &  33 & $>$308 & 0.25 &     \\
$12 \rightarrow 10$     &        & 191 & $148^{+66}_{-148}$ &  49  & 853 \\
$14 \rightarrow 12$     &        &  16 &        &  70  & 273 \\
$14_3 \rightarrow 12$   &        &  32 &        &  42  &     \\
$16 \rightarrow 14$     &        & 219 &        & 199  & 672 \\
$16_4 \rightarrow 14_3$ &        & 151 &        & 107  &     \\ \hline
$2_2 \rightarrow 0$     & $1.6 \pm 0.6$ &   4 & $2.3 \pm 0.4$ & 0.5  &     \\
$2_2 \rightarrow 2$     & $269 \pm 127$ & 233 & $8.0 \pm 3.5$ & 375  &     \\ \hline
\end{tabular}
\label{table1}
\end{table}
%=========================================================

   There are not sufficient data on $B(E2)$ for $^{66}$Ge.  The experimental
large value of $B(E2:2_2^+ \rightarrow 2_1^+)$ and small value of
$B(E2:2_2^+ \rightarrow 0_1^+)$ are well reproduced by the calculations.
Although the calculated value 281 $e^2$fm$^{4}$ of $B(E2:2_1^+ \rightarrow 0_1^+)$
is larger than the experimental one $190 \pm 36$ $e^2$fm$^{4}$, the qualitative
trend of the observed $\Delta J =2$ transitions \cite{Stefanova} 
in the two cascade bands (the $gs$ band and the second positive-parity band
on $2_2^+$) up to $J=8$ can be explained by the calculated $B(E2)$ values. 
The large ratio of $B(E2:2_2^+ \rightarrow 2_1^+)/B(E2:2_2^+ \rightarrow 0_1^+)$ 
was discussed as a signature of the triaxiality of $^{64}$Ge in Ref.
\cite{Kaneko}.  The experimental and calculated ratios suggest the same nature
of $^{66}$Ge as that of $^{64}$Ge, which is consistent with the previous
understanding \cite{Stefanova} of the $\gamma$ softness.
There are similarities between $^{66}$Ge and $^{64}$Ge with respect to other
$B(E2)$ values calculated with the parameters (2), (3), and (4).
We can presume a similar structure in the low-spin states
before the backbending in $^{66}$Ge and $^{64}$Ge.

  It is notable in the calculated results for $^{66}$Ge
that $B(E2:10_1^+ \rightarrow 8_1^+)$ is small in correspondence to
the backbending at $10_1^+$ seen in Fig. \ref{fig4}(a).
This shows that the structure of the yrast states changes at $10_1^+$
in $^{66}$Ge.  The larger $B(E2:10_1^+ \rightarrow 8_2^+)$ value
than $B(E2:10_1^+ \rightarrow 8_1^+)$ in the calculation (probably in experiment)
shows the continuation of the $10_1^+$ state to the $8_2^+$ state.
 These results indicate a band crossing between $J=8$ and $J=10$.
The $10_2^+$ state is, however, not connected to the $8_1^+$ state, 
because $B(E2:10_2^+ \rightarrow 8_1^+)$ is small in theory.
The experimentally observed inter-band transition $8_2^+ \rightarrow 8_1^+$
suggests significant mixing of $8^+$ states.
  Above $10_1^+$, the calculated $B(E2:J_1^+ \rightarrow (J-2)_1^+)$ values
between the yrast states become quite large up to $18_1^+$,
except that $B(E2:14_1^+ \rightarrow 12_1^+)$ is small.
This explains the cascade of $\Delta J=2$ transitions from the tentative $16^+$
state observed in Ref. \cite{Stefanova}.  The small value of
$B(E2:14_1^+ \rightarrow 12_1^+)$ corresponds to the structural change
at $14_1^+$ suggested by the $J - \omega$ graph in Fig. \ref{fig4}(a).
   Our model predicts another interesting backbending at $11_1^+$
in the sequence of the odd-$J$ positive-parity yrast states.  As shown
in Table \ref{table2}, the calculated $B(E2:11_1^+ \rightarrow 9_1^+)$ value
is very small, while $B(E2:\Delta J=2)$ is large below $9_1^+$ and
above $11_1^+$.
This shows that the structure changes at the $11_1^+$ state. 
In other words, the sequence of the odd-$J$ positive-parity yrast states
is composed of two bands, from $3_1^+$ till $9_1^+$ and from $11_1^+$ to $19^+$.
Also for the sequence of the odd-$J$ negative-parity yrast states of $^{66}$Ge,
the $B(E2:15_1^- \rightarrow 13_1^-)$ value is very small, which corresponds
to the backbending at $15_1^-$ both in the experimental and calculated
$J -\omega$ graphs shown in Fig. \ref{fig4}(b).  

   The experiment for $^{66}$Ge \cite{Stefanova} found a cascade
of $\Delta J=1$ transitions connecting two $\Delta J=2$ sequences
with even $J$ and odd $J$
($15_1^+ \rightarrow 14_1^+$ and
 $13_1^+ \rightarrow 12_1^+ \rightarrow 11_1^+ \rightarrow 10_1^+$),
 and supposed that the $\Delta J=1$ transitions are $M1$ transitions.
We calculated both of $B(E2)$ and $B(M1)$ to examine this supposition.
The results are shown in Table \ref{table2}. 
 The $B(M1)$ values obtained with our model are consistent with the results 
of Ref. \cite{Stefanova} that $B(M1:12_1^+ \rightarrow 11_1^+)$ is large,
and $B(M1:13_1^+ \rightarrow 12_1^+)$ and $B(M1:15_1^+ \rightarrow 14_1^+)$
are small.  However, the $B(M1:14_1^+ \rightarrow 13_1^+)$ value is small
in contrast to the predicted staggering.
 The experimental value of $B(M1:2_2^+ \rightarrow 2_1^+)$ is 0.008 $\mu _N^2$
and the calculated one is 0.0023 $\mu _N^2$. The roughly good prediction
testifies to a considerable reliability of our $B(M1)$ values.
Table \ref{table2} shows that the $B(E2:\Delta J=1)$ 
values are not negligible for the states $10_1^+$ to $15_1^+$,
which suggests possible mixing of $E2$ and $M1$ transitions.

%===============  table 2  ========================
\begin{table}[b]
\caption{Calculated $B(E2)$ ($e^2$fm$^{4}$) and $B(M1)$ ($\mu _N^2$) values
         for the positive-parity states from $J=9$ to $J=15$
         of $^{66}$Ge and $^{68}$Ge.
         Non-yrast states are distinguished with their subscripts
         from the yrast states with no subscript.}
\begin{tabular}{l|c|c||c|c}   \hline
        & \multicolumn{2}{|c||}{$^{66}$Ge cal}
         & \multicolumn{2}{c}{$^{68}$Ge cal}  \\ \hline
$J_i^+ \rightarrow J_f^+$ & $B(E2)$ & $B(M1)$ & $B(E2)$ & $B(M1)$  \\ \hline
$11 \rightarrow 9$     & 0.14 &       & 1.2  &        \\
$11 \rightarrow 10$    &  11  & 0.036 &   7  & 0.001  \\
$11 \rightarrow 10_2$  &  16  & 0.26  &  22  & 0.33  \\
$12 \rightarrow 10$    & 191  &       &  49  &        \\
$12 \rightarrow 11$    &  22  & 0.66  & 0.5  & 0.14   \\
$13 \rightarrow 11$    & 312  &       & 180  &        \\
$13 \rightarrow 12$    &  17  & 0.021 &  19  & 0.030  \\
$14 \rightarrow 12$    &  16  &       &  70  &        \\
$14 \rightarrow 13$    &  23  & 0.003 &   9  & 0.043  \\
$14_3 \rightarrow 13$  &  0.4 &       & 0.02 & 0.044  \\
$15 \rightarrow 13$    & 270  &       & 166  &        \\
$15 \rightarrow 14$    &   3  & 0.015 &   4  & 0.20   \\
$15 \rightarrow 14_3$  &  12  &       &   6  & 0.048  \\ \hline
\end{tabular}
\label{table2}
\end{table}
%=========================================================

   For $^{68}$Ge, experimental $B(E2)$ values of the $gs$  band up to $6_1^+$
are better reproduced by the present shell model calculation as compared with
the VAMPIR calculation as shown in Table \ref{table2}.
 Except for several band crossings, our model successfully predicts
large $E2$ transitions in the four sequences of the positive- and
 negative-parity states with even $J$ and odd $J$
deduced by cascades of $\Delta J=2$ transitions in Ref. \cite{Ward}.
Some of the calculated $B(E2)$ values for $8^+ \rightarrow 6^+$ transitions
do not correspond with the experimental values.
The small value of $B(E2:8_1^+ \rightarrow 6_1^+)$ is notable
in the calculated result.  
This is in agreement with the backbending at $8_1^+$
in Fig. \ref{fig5}(a), which suggests that a structural change takes place
at the $8_1^+$ state in $^{68}$Ge, in contrast to the backbending
at $10_1^+$ and the small value of $B(E2:10_1^+ \rightarrow 8_1^+)$
in $^{66}$Ge.  
  The calculation predicts a large value for $B(E2:8_2^+ \rightarrow 6_1^+)$,
suggesting that the $gs$ band up to $6_1^+$ is connected to the $8_2^+$ state,
and a band crossing happens between $J=6$ and $J=8$. 
  This is consistent with the result of the VAMPIR calculation,
but contradicts the result of the particle-rotor model \cite{Lima}
in which the $8_3^+$ state is identified as the continuation
of the $gs$ band. 
 We note again that the sequence of the states ($8_1^+$, $10_1^+$, $12_1^+$,
$14_4^+$, $16_6^+$) connected by strong $E2$ transitions in experiment
corresponds well with the sequence of the states
($8_1^+$, $10_1^+$, $12_1^+$, $14_3^+$, $16_4^+$) connected by large $B(E2)$
values in theory.
To other bands assigned in the experiment \cite{Ward}, our model
has corresponding states connected by large $B(E2)$ values
except that $B(E2)$ values become small at band crossings.

   The $11_1^+$ state has not experimentally been observed in the sequence
of the odd-$J$ positive-parity yrast states of $^{68}$Ge.  Our model provides
very small values for $B(E2:11_1^+ \rightarrow 9_1^+)$ and
$B(E2:9_1^+ \rightarrow 7_1^+)$, which suggests a structural change near
the $9_1^+$ and $11_1^+$ states. The situation is rather complicated
as compared with the backebnding in other bands.
   In order to compare the $M1$ transitions between the states from $J=10$ to
$J=15$ in $^{68}$Ge with those in $^{66}$Ge, we calculated $B(M1)$ and $B(E2)$
values also for $^{68}$Ge.  Table \ref{table2} shows that the calculated
$B(M1)$ values for these states of $^{68}$Ge are, up to $14^+$, similar to
those of $^{66}$Ge but $B(M1:15_1^+ \rightarrow 14_1^+)$ is large in $^{68}$Ge
in contrast to $^{66}$Ge.
For these states, the $B(E2:\Delta J=1)$ values are a little smaller
in $^{68}$Ge than in $^{66}$Ge.
 Our model suggests that the $M1$ transition could contribute
to the $\Delta J=1$ transitions also in $^{68}$Ge.

   For $^{68}$Ge, the continuation of the negative-parity states ($17_1^-$,
$19_1^-$, $21_1^-$, $23_1^-$) and the termination of the odd-$J$
negative-parity band are discussed in Ref. \cite{Ward}.
The present shell model, which reproduces well these experimental energy levels
as shown in Fig. \ref{fig2}, yields the following $B(E2)$ values: 156, 234,
218, and 243 in $e^2$fm$^4$ for the transitions
 $23_1^- \rightarrow 21_2^- \rightarrow 19_2^- \rightarrow
  17_1^- \rightarrow 15_1^-$; 0.1, 163, and 9 in $e^2$fm$^4$ for the transitions
 $23_1^- \rightarrow 21_1^- \rightarrow 19_1^- \rightarrow 17_1^-$.
These $B(E2)$ values indicate the continuation of the states
 $17_1^-$, $19_2^-$, $21_2^-$, and $23_1^-$.
This result disagrees with the considerations in Ref. \cite{Ward},
where the $17_1^-$ state is considered to be a possible terminating state.
  Also in $^{66}$Ge, as shown in Fig. \ref{fig1}, the calculated $B(E2)$ values
suggest the continuation of the states $17_1^-$, $19_2^-$, and $21_2^-$.
The calculations, however, show a difference between $^{68}$Ge and $^{66}$Ge,
i.e., $B(E2:23_1^+ \rightarrow 21_2^+)$ is very small as compared with
$B(E2:23_1^+ \rightarrow 21_1^+)$ in $^{66}$Ge in opposition to $^{68}$Ge.
This remains as a question about the band on $15_1^-$ in $^{66}$Ge.

   We have seen backbending at $15_1^-$ in the sequence of the odd-$J$
negative-parity yrast states of $^{68}$Ge both in experiment and theory,
in Fig. \ref{fig5}(b).
 The small value of $B(E2:15_1^- \rightarrow 13_1^-)$ indicates a change
in the structure of this sequence of states.  The backbending at $15_1^-$
corresponds to that in $^{66}$Ge.
The calculated result in Fig. \ref{fig2} shows a sign of backbending
at $14_1^-$ in the even-$J$ negative-parity yrast states.
 However, the small values of $B(E2:14_1^- \rightarrow 12_1^-)$ and 
$B(E2:12_1^- \rightarrow 10_1^-)$ suggest a complicated structural change
near the $12_1^-$ and $14_1^-$ states which is similar to that of the odd-$J$ 
positive-parity yrast states.  The complicated situations may be related to
the missing of $12_1^-$ and $11_1^+$ in the experiment \cite{Ward}.

%=================================================================
\section{Particle alignments}\label{sec4}

\subsection{Particle alignments in the even-$J$ positive-parity yrast states}\label{sec4.1}

%===============  table 3  ========================
\begin{table}[b]
\caption{Expectation values of proton and neutron numbers
         in the four orbits, calculated for the even-$J$ positive-parity
         yrast states and some low-energy states of $^{66}$Ge.
         Calculated $Q$ moments (in $e$ fm$^2$) are also tabulated.}
\begin{tabular}{c|cccc|cccc|c}   \hline
        & \multicolumn{4}{c}{proton} & \multicolumn{4}{|c|}{neutron} & \\ \hline
$^{66}$Ge & $p_{3/2}$ & $f_{5/2}$ & $p_{1/2}$ & $g_{9/2}$
         & $p_{3/2}$ & $f_{5/2}$ & $p_{1/2}$ & $g_{9/2}$ & $Q$  \\ \hline
  $0_1^+$  & 1.72 & 1.72 & 0.50 & 0.06 & 2.42 & 2.68 & 0.77 & 0.13 &       \\
  $2_1^+$  & 1.71 & 1.70 & 0.53 & 0.06 & 2.40 & 2.77 & 0.70 & 0.13 & -23.2 \\
  $4_1^+$  & 1.72 & 1.72 & 0.49 & 0.07 & 2.39 & 2.82 & 0.67 & 0.12 & -27.9 \\
  $6_1^+$  & 1.71 & 1.79 & 0.43 & 0.07 & 2.35 & 2.83 & 0.70 & 0.12 & -35.8 \\
  $8_1^+$  & 1.67 & 1.84 & 0.42 & 0.08 & 2.42 & 2.67 & 0.74 & 0.16 & -39.4 \\ \hline
  $8_2^+$  & 1.60 & 1.72 & 0.51 & 0.18 & 1.62 & 1.88 & 0.61 & 1.89 & -59.9 \\
 $10_1^+$  & 1.45 & 1.49 & 0.63 & 0.43 & 1.75 & 1.95 & 0.61 & 1.69 & -70.8 \\
 $12_4^+$  & 1.52 & 1.53 & 0.54 & 0.41 & 1.69 & 2.02 & 0.61 & 1.67 & -58.5 \\ \hline
 $10_4^+$  & 1.19 & 1.20 & 0.64 & 0.98 & 1.78 & 2.20 & 0.94 & 1.08 & -80.2 \\
 $12_1^+$  & 1.30 & 1.17 & 0.61 & 0.93 & 2.05 & 2.16 & 0.59 & 1.20 & -80.5 \\
 $14_1^+$  & 1.29 & 1.17 & 0.54 & 1.00 & 1.97 & 2.12 & 0.85 & 1.06 & -83.7 \\
 $16_1^+$  & 1.28 & 1.15 & 0.57 & 1.00 & 1.88 & 2.38 & 0.66 & 1.08 & -80.5 \\
 $18_1^+$  & 1.26 & 1.17 & 0.55 & 1.02 & 1.84 & 2.45 & 0.63 & 1.07 & -81.8 \\ \hline
 $18_2^+$  & 0.64 & 0.76 & 0.61 & 1.99 & 1.59 & 1.78 & 0.61 & 2.03 & -83.9 \\
 $20_1^+$  & 0.73 & 0.84 & 0.48 & 1.95 & 1.55 & 1.73 & 0.64 & 2.07 & -84.7 \\
 $22_1^+$  & 0.69 & 0.84 & 0.46 & 2.00 & 1.38 & 2.02 & 0.59 & 2.01 & -82.9 \\
 $24_1^+$  & 0.78 & 1.03 & 0.19 & 2.00 & 1.38 & 2.07 & 0.54 & 2.01 & -84.7 \\
 $26_1^+$  & 0.91 & 1.09 & 0.00 & 2.01 & 1.32 & 2.09 & 0.58 & 2.02 & -86.1 \\ \hline
  $2_2^+$  & 1.69 & 1.75 & 0.48 & 0.09 & 2.38 & 2.55 & 0.93 & 0.14 & +20.9 \\
  $4_2^+$  & 1.75 & 1.61 & 0.57 & 0.08 & 2.40 & 2.78 & 0.68 & 0.14 & -20.6 \\
  $6_2^+$  & 1.70 & 1.70 & 0.52 & 0.09 & 2.48 & 2.72 & 0.68 & 0.14 & -16.5 \\ \hline
  $3_1^+$  & 1.79 & 1.56 & 0.56 & 0.10 & 2.44 & 2.72 & 0.70 & 0.14 &  -0.9 \\
  $4_3^+$  & 1.69 & 1.62 & 0.60 & 0.09 & 2.50 & 2.69 & 0.67 & 0.14 &  +5.6 \\ \hline
\end{tabular}
\label{table3}
\end{table}
%=========================================================

\begin{figure}
\includegraphics[width=6.8cm,height=6.8cm]{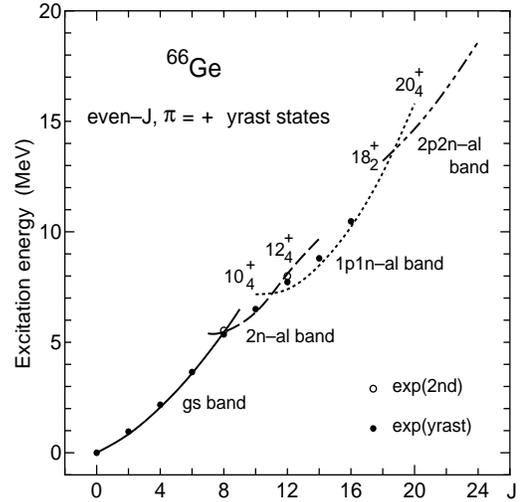}
  \caption{Comparison of the calculated four bands (curves)
           with the experimental yrast states (solid circles) for $^{66}$Ge.}
  \label{fig6}
\end{figure}

   We calculated expectation values of proton and neutron numbers in each orbit
($\langle n_a^\pi \rangle$ and $\langle n_a^\nu \rangle$), and the spectroscopic
$Q$ moment, in order to investigate the structure of $^{66}$Ge and $^{68}$Ge.
The calculated results for the even-$J$ positive-parity yrast states
and some other states of $^{66}$Ge are shown in Table \ref{table3}.
  We have shown in a recent paper \cite{Hase4} that the structural changes
  revealed in the $J-\omega$ graph of Fig. \ref{fig4}(a)
 can be explained by successive alignments of three combinations of nucleons
 in the $g_{9/2}$  orbit: two-neutron ($2n$) alignment coupled to $T=1$, $J=8$;
 one-proton-one-neutron ($1p1n$) alignment coupled to $T=0$, $J=9$;
 two-proton-two-neutron ($2p2n$) alignment coupled to $T=0$, $J=16$, 
 i.e., $[(g_{9/2}^\pi)^2_{T=1,J=8}(g_{9/2}^\nu)^2_{T=1,J=8}]_{T=0,J=16}$.
 The scenario of the changes is as follows: (a) The $2n$ alignment takes place
 at the $8_2^+$ state.  The $2n$ aligned band crosses the $gs$ band between
 $J=8$ and $J=10$. (b) The $1p1n$ aligned band competes with the $2n$
 aligned band near $J=10$ and $J=12$, and there is interplay between
 the two bands. (c) The $1p1n$ alignment overwhelms the $2n$ alignment
 at $14_1^+$, and the $1p1n$ aligned band appears in the yrast line
 from $14_1^+$ to $18_1^+$. (d) The $2p2n$ aligned band takes over as the
 yrast state at $20_1^+$, continuing up to the $26_1^+$ state
 where the band terminates.  This scenario is narrated by the change
 of the proton and neutron numbers listed in Table \ref{table3}.
 The successive alternations of the four bands cause the changes of the
 $B(E2)$ values at the band crossings, as shown in section~\ref{sec3}.
  The abrupt change of the spectroscopic $Q$ moment from $8_1^+$ to $10_1^+$
 also testifies to the band crossing.
 We illustrate the successive four bands with a graph of the excitation energy
 versus the spin $J$ (we call it the ``$E_x -J$ graph"), in Fig. \ref{fig6}.
 Here, the $10_4^+$ state has a dominant component of the aligned $1p1n$ pair
 in the $g_{9/2}$ orbit, the $12_4^+$ state has a dominant component of
 the aligned $2n$ pair in the $g_{9/2}$ orbit, and the $18_2^+$ state
 belongs to the $2p2n$ aligned band.

%===============  fig. 6  ========================================
%=================================================================

%===============  table 4  ========================
\begin{table}[h]
\caption{Expectation values of proton and neutron numbers
         in the four orbits, calculated for the even-$J$ positive-parity
         yrast states and some low-energy states of $^{68}$Ge.
         Calculated $Q$ moments (in $e$ fm$^2$) are also tabulated.}
\begin{tabular}{c|cccc|cccc|c}   \hline
        & \multicolumn{4}{c}{proton} & \multicolumn{4}{|c|}{neutron} & \\ \hline
$^{68}$Ge & $p_{3/2}$ & $f_{5/2}$ & $p_{1/2}$ & $g_{9/2}$
         & $p_{3/2}$ & $f_{5/2}$ & $p_{1/2}$ & $g_{9/2}$ & $Q$  \\ \hline
  $0_1^+$  & 1.74 & 1.69 & 0.51 & 0.07 & 3.05 & 3.50 & 1.21 & 0.24 &       \\
  $2_1^+$  & 1.72 & 1.69 & 0.51 & 0.08 & 3.02 & 3.51 & 1.24 & 0.23 &  -6.8 \\
  $4_1^+$  & 1.75 & 1.70 & 0.46 & 0.09 & 3.02 & 3.46 & 1.31 & 0.22 &  -8.1 \\
  $6_1^+$  & 1.76 & 1.73 & 0.40 & 0.11 & 3.02 & 3.39 & 1.37 & 0.22 & -13.3 \\
  $8_2^+$  & 1.80 & 1.79 & 0.29 & 0.12 & 2.98 & 3.22 & 1.54 & 0.25 & -13.2 \\ \hline
  $8_1^+$  & 1.66 & 1.72 & 0.50 & 0.12 & 2.22 & 2.85 & 0.92 & 2.02 & -63.5 \\
 $10_1^+$  & 1.61 & 1.60 & 0.64 & 0.16 & 2.22 & 2.93 & 0.85 & 1.99 & -69.8 \\
 $12_3^+$  & 1.62 & 1.66 & 0.56 & 0.16 & 2.22 & 3.04 & 0.77 & 1.97 & -60.4 \\
 $14_3^+$  & 1.38 & 1.47 & 0.58 & 0.57 & 2.47 & 2.94 & 1.02 & 1.57 & -77.0 \\
 $16_4^+$  & 1.51 & 1.85 & 0.52 & 0.12 & 2.26 & 2.83 & 0.90 & 2.02 & -62.3 \\ \hline
 $12_1^+$  & 1.26 & 1.22 & 0.75 & 0.78 & 2.57 & 2.86 & 1.15 & 1.42 & -87.8 \\
 $14_1^+$  & 1.27 & 1.17 & 0.59 & 0.98 & 2.75 & 3.02 & 1.08 & 1.16 & -88.6 \\
 $16_1^+$  & 1.28 & 1.19 & 0.61 & 0.98 & 2.82 & 2.82 & 1.20 & 1.17 & -91.6 \\
 $18_1^+$  & 1.20 & 1.17 & 0.61 & 1.03 & 2.89 & 2.76 & 1.24 & 1.11 & -94.3 \\ \hline
 $18_2^+$  & 0.67 & 0.73 & 0.65 & 1.95 & 2.19 & 3.03 & 0.69 & 2.10 & -83.3 \\
 $20_1^+$  & 0.75 & 0.84 & 0.51 & 1.90 & 2.20 & 2.89 & 0.76 & 2.15 & -86.2 \\
 $22_1^+$  & 0.70 & 0.90 & 0.46 & 1.94 & 2.34 & 2.70 & 0.86 & 2.11 & -88.6 \\
 $24_1^+$  & 0.87 & 1.12 & 0.01 & 2.00 & 2.41 & 2.33 & 1.23 & 2.03 & -93.9 \\
 $26_1^+$  & 0.88 & 1.12 & 0.00 & 2.00 & 2.09 & 2.64 & 1.21 & 2.06 & -89.0 \\ \hline
  $0_2^+$  & 1.72 & 1.68 & 0.49 & 0.11 & 2.52 & 3.07 & 1.30 & 1.12 &       \\
  $2_2^+$  & 1.71 & 1.63 & 0.57 & 0.09 & 2.98 & 3.57 & 1.22 & 0.22 &  +7.8 \\
  $4_3^+$  & 1.75 & 1.61 & 0.54 & 0.10 & 3.14 & 3.71 & 0.93 & 0.22 & +41.8 \\
  $6_2^+$  & 1.72 & 1.61 & 0.56 & 0.11 & 3.05 & 3.59 & 1.15 & 0.21 & +12.0 \\ \hline
  $4_2^+$  & 1.67 & 1.34 & 0.90 & 0.09 & 3.05 & 3.55 & 1.19 & 0.21 & -42.8 \\
  $8_3^+$  & 1.62 & 1.69 & 0.49 & 0.20 & 2.24 & 2.80 & 1.00 & 1.96 & -59.2 \\
 $10_2^+$  & 1.41 & 1.41 & 0.65 & 0.54 & 2.34 & 3.02 & 1.01 & 1.64 & -76.1 \\
 $12_2^+$  & 1.29 & 1.22 & 0.69 & 0.80 & 2.52 & 3.10 & 1.01 & 1.38 & -85.1 \\ \hline
\end{tabular}
\label{table4}
\end{table}
%=========================================================

\begin{figure}[h]
\includegraphics[width=6.0cm,height=6.0cm]{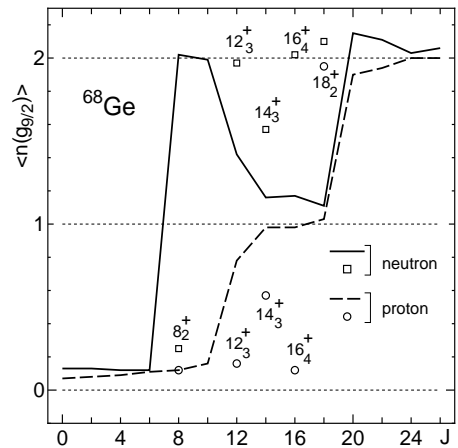}
  \caption{The expectation values of neutron and proton numbers
           in the $g_{9/2}$ orbit ($\langle n^\nu _{g9/2} \rangle$
            and $\langle n^\pi _{g9/2} \rangle$)
            for the even-$J$ positive-parity yrast states (lines)
            and some other states (marks) of $^{68}$Ge.}
  \label{fig7}
\end{figure}

\begin{figure}[b]
\includegraphics[width=6.0cm,height=8.0cm]{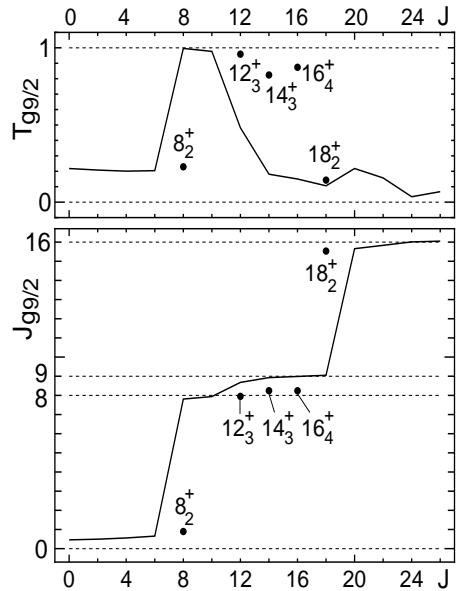}
   \caption{The expectation values of the spin and isospin of nucleons
            in the $g_{9/2}$ orbit ($J_{g9/2}$ and $T_{g9/2}$)
            for the even-$J$ positive-parity yrast states (lines)
            and some other states (marks) of $^{68}$Ge.}
  \label{fig8}
\end{figure}

   Let us show the results for the even-$J$ positive-parity yrast states of
$^{68}$Ge, instead of repeating the results for $^{66}$Ge shown in
Ref. \cite{Hase4}.  Table \ref{table4} lists the proton and neutron
numbers, $\langle n_a^\pi \rangle$ and $\langle n_a^\nu \rangle$, for the even-$J$
positive-parity yrast states and some other states of $^{68}$Ge.
This table shows that the structural changes parallel to those mentioned above
for $^{66}$Ge happen in $^{68}$Ge.
A remarkable difference of $^{68}$Ge from $^{66}$Ge is the $2n$ alignment
at the $8_1^+$ state of $^{68}$Ge, where the neutron number 
$\langle n_{g9/2}^\nu \rangle \approx 2$ indicates the $2n$ alignment
in the $g_{9/2}$ orbit.  We illustrate the changes of
$\langle n_{g9/2}^\pi \rangle$ and $\langle n_{g9/2}^\nu \rangle$
in Fig. \ref{fig7}, and expectation values of the spin and isospin of
nucleons in the $g_{9/2}$ orbit (which are denoted by $J_{g9/2}$ and $T_{g9/2}$)
in Fig. \ref{fig8}.    
Here, we evaluate the spin $J_{g9/2}$ and the isospin $T_{g9/2}$ as follows:
$J_{g9/2}=[\langle ({\hat j}_{g9/2})^2 \rangle +1/4]^{1/2} -1/2$ and
$T_{g9/2}=[\langle ({\hat t}_{g9/2})^2 \rangle +1/4]^{1/2} -1/2$.
Figures \ref{fig7} and \ref{fig8} narrate the same scenario as that
mentioned above for $^{66}$Ge.

%===============  fig. 7 ========================================
%=================================================================

%===============  fig. 8  ========================================
%=================================================================

%===============  fig. 9  ========================================
\begin{figure}[t]
\includegraphics[width=7.0cm,height=7.0cm]{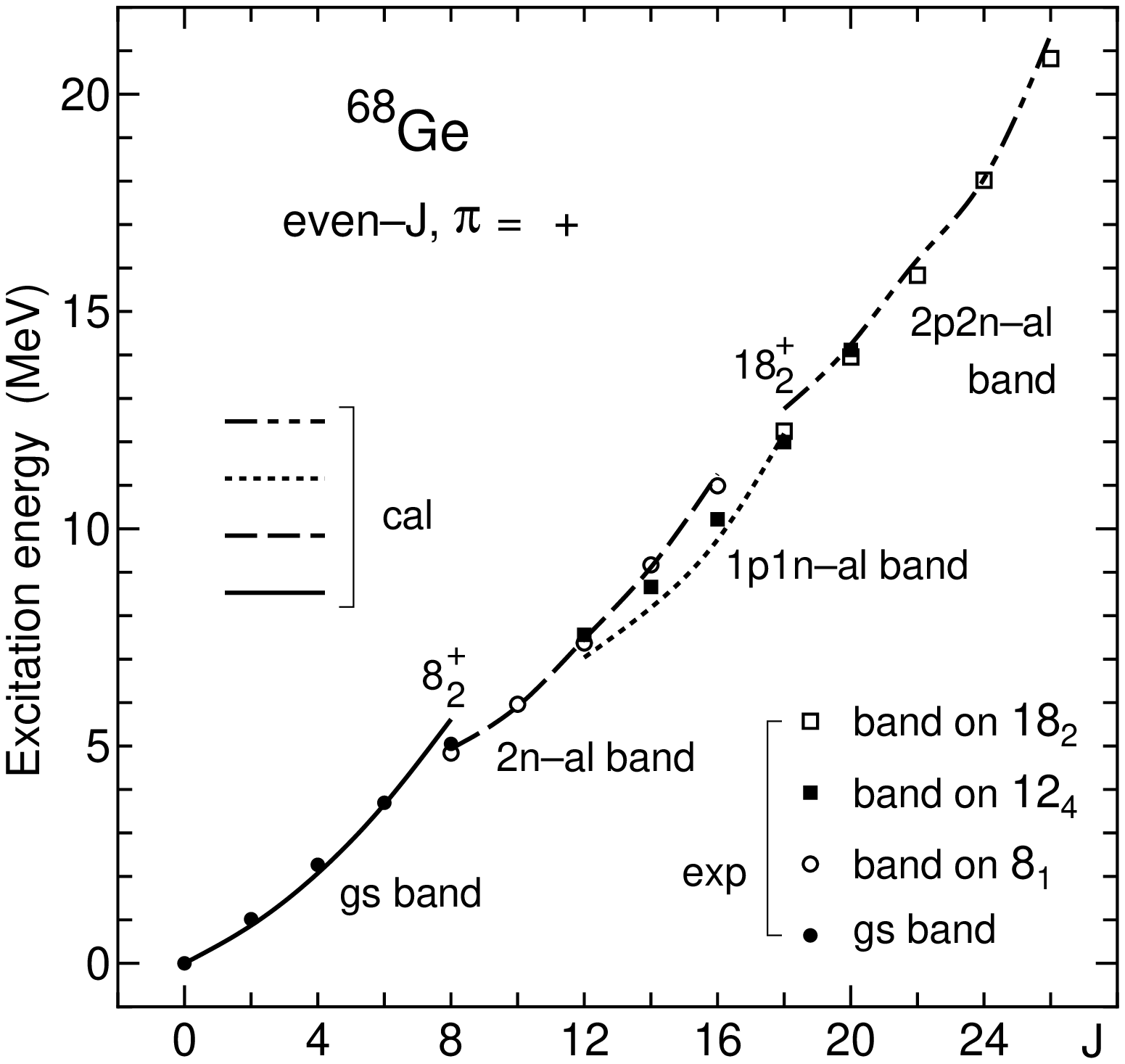}
  \caption{Comparison of the calculated four bands with the experimentally
           observed bands for $^{68}$Ge.}
  \label{fig9}
\end{figure}
%=================================================================

   In $^{68}$Ge, while the $gs$ band continues to the $8_2^+$ state,
the $2n$ aligned band crosses the $gs$ band before $J=8$. 
The abrupt change of the $Q$ moment at $8_1^+$ indicates the band crossing.
 The yrast states $8_1^+$ and $10_1^+$ are members of the $2n$ aligned band.
Passing the transitional state $12_1^+$, the $1p1n$ aligned band appears
in the yrast line of $14_1^+$, $16_1^+$, and $18_1^+$ like $^{66}$Ge.
There seems to be strong coupling between the $2n$ and $1p1n$ aligned bands
at $J=12$.  The third band crossing takes place between $J=18$ and $J=20$.
The $2p2n$ aligned band takes the lowest position from $20_1^+$ to $26_1^+$.
Figure \ref{fig8} shows that the aligned nucleons in the $g_{9/2}$ orbit
couple to $T=1$, $J=8$ in the $2n$ aligned band ($8_1^+$, $10_1^+$),
$T=0$, $J=9$ in the $1p1n$ aligned band ($14_1^+$, $16_1^+$, $18_1^+$)
and $T=0$, $J=16$ in the $2p2n$ aligned band
($20_1^+$, $22_1^+$, $\cdots$, $26_1^+$).
We show the successive four bands in the $E_x -J$ graph, in Fig. \ref{fig9},
where corresponding experimental bands are taken from Ref. \cite{Ward}.
The agreement between theory and experiment is satisfactory.
It should be noted that the correspondence between theory and experiment
is good for the non-yrast states as well as the yrast states.

   However, the $12_1^+$ state is assigned as the $J=12$ member of the
band on $8_1^+$ in the experiment \cite{Ward}, while the $12_1^+$ state
resembles the $1p1n$ aligned state rather than the $2n$ aligned state
in our calculation.  The $J-\omega$ graph for the even-$J$ positive-parity
yrast states of $^{68}$Ge (Fig. \ref{fig5}(a)) shows
an insufficient description of the energy difference $E(18_1^+)-E(16_1^+)$.
The calculated $B(E2)$ values between different bands do not agree
with the experimental $B(E2)$ values in Table \ref{table1}, either.
These discrepancies reveal missing correlations in our model
Hamiltonian, and suggest stronger interplays between the different bands
near the band crossings.  Still, the calculated $J-\omega$ graph
in Fig. \ref{fig5}(a) and the $E_x -J$ graph in Fig. \ref{fig9} trace the trends
of the experimental graphs.  We can attribute the changes in these graphs
to the successive band crossings indicated in Figs. \ref{fig7} and \ref{fig8}.

\begin{figure}[t]
\includegraphics[width=7.0cm,height=8.0cm]{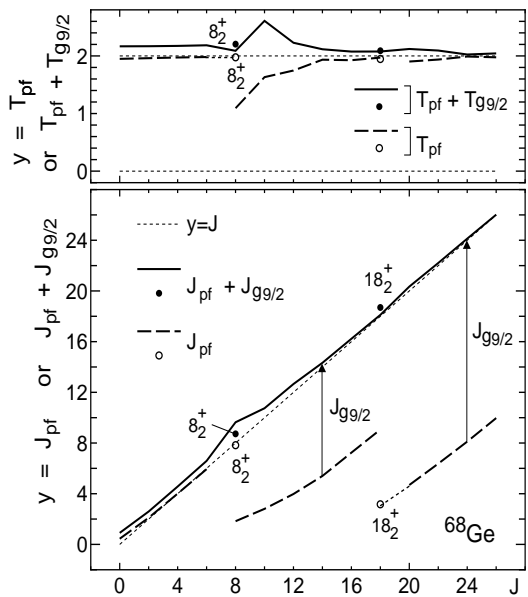}
   \caption{The expectation values of spins $J_{pf}$ and $J_{g9/2}$,
            and those of isospins $T_{pf}$ and $T_{g9/2}$,
            for the even-$J$ positive-parity yrast states (lines)
            and some other states (marks) of $^{68}$Ge.
            }
  \label{fig10}
\end{figure}

   So far, we have used the word ``alignment" for the maximum angular momentum
coupling in the $g_{9/2}$ orbit.  How is the angular momentum coupling
of the aligned particles in the $g_{9/2}$ orbit with the central system
excluding the $g_{9/2}$ particles?
We calculated expectation values of the spin and isospin of the central
system which is represented by nucleons in the $pf$ shell
($2p_{3/2}$, $1f_{5/2}$, $2p_{1/2}$) in our shell model.
We write these expectation values as $J_{pf}$ and $T_{pf}$, and evaluate
them using the relations 
$J_{pf}=[\langle ({\hat j}_{pf})^2 \rangle +1/4]^{1/2} -1/2$ and
$T_{pf}=[\langle ({\hat t}_{pf})^2 \rangle +1/4]^{1/2} -1/2$,
where ${\hat j}_{pf}$ and ${\hat t}_{pf}$ mean the spin and isospin operators
for the subspace ($2p_{3/2}$, $1f_{5/2}$, $2p_{1/2}$).
Calculated $J_{pf}$ and $T_{pf}$ together with $J_{g9/2}$ and $T_{g9/2}$ 
are illustrated in Fig. \ref{fig10}.
Figure \ref{fig10} indicates that the spin of the aligned particles
in the $g_{9/2}$ orbit actually aligns with the spin of the central system
in the $2n$, $1p1n$ and $2p2n$ aligned bands.
This situation can be called a system composed of a rotor and particles.
We can see some deviation from the weak coupling of the rotor and particles
near $J=8$ in Fig. \ref{fig10}.
The upper panel of Fig. \ref{fig10} indicates vector coupling of the isospins 
$T_{pf}$ and $T_{g9/2}$ for the $2n$ aligned yrast states with $T=1$,
$8_1^+$ and $10_1^+$.
  It should be noticed that the residual nucleons in the $pf$ shell
coupled with the aligned $1p1n$ pair with $T=0$, $J=9$ must have
the isospin $T=1$ for the nucleus $^{66}$Ge, while the residual nucleons
coupled with the aligned $2n$ pair with $T=1$, $J=8$ can have the isospins
$T=0$ and $T=1$.  This effect is seen in the upper panel of Fig. \ref{fig10}.
The different isospin couplings bring about different properties to
the $1p1n$ and $2n$ aligned bands.  The problem is related to the competition
between the $T=1$ and $T=0$ pair correlations in the central system which is
represented by the $pf$-shell nucleons in our shell model.

  The superiority of the $J=9$, $T=0$ $1p1n$ pair in the $14_1^+$, $16_1^+$,
 and $18_1^+$ states can be attributed to the condition that the $T=0$, $J=9$
 $pn$ interaction is stronger than the $T=1$, $J=8$ interaction.
 Note that while the $T=1$, $J=2j-1$ interaction is repulsive, the $T=0$,
 $J=2j$ interaction is very attractive in ordinary effective interactions.
 If we set $ \langle (g_{9/2})^2 |V|(g_{9/2})^2:T=0,J=9 \rangle $ zero,
 the $1p1n$ aligned states do not become the yrast states, while the $gs$ band
 is hardly disturbed.

%===============  fig. 10  ========================================
%=================================================================

   What conditions cause such a nearly pure $1p1n$  alignment?
In Ref. \cite{Hase3}, we investigated even-mass Ru isotopes around $^{90}$Ru
which is symmetrical to $^{66}$Ge with respect to the particle-hole
transformation in the ($p_{3/2}$,$f_{5/2}$,$p_{1/2}$,$g_{9/2}$) space.
We did not find any sign of the $T=0$ $1p1n$ alignment there,
and could not see a pure $2n$ alignment at the backbending state $8_1^+$
in $^{90}$Ru.
% We observed $\alpha$-like correlations and interplays between
%the quartets and aligned nucleons in the $g_{9/2}$ orbit in the Ru isotopes.
An important thing is that the Fermi level lies at the $g_{9/2}$ orbit itself
in the Ru isotopes but considerably far from the $g_{9/2}$ orbit
in the Ge isotopes.
 The appearance of the nearly pure $2n$ and $1p1n$ alignments in $^{66}$Ge
is based on the condition that the high-spin orbit $g_{9/2}$ is
quite apart from the Fermi level and has the opposite parity to the $pf$ shell.
 Only even-number nucleons are allowed to occupy the $g_{9/2}$ orbit
 after covering the cost of excitation energy from $pf$ to $g_{9/2}$.
 We can expect the $T=0$ $1p1n$ alignment in $N \approx Z$ even-even nuclei
 near the Ge isotopes. 

  Carrying out the same shell model calculation, we explored other nuclei
$^{64}$Ge, $^{70}$Ge, $^{60-68}$Zn, and $^{68}$Se for the $T=0$ $1p1n$
alignment.  The calculation for $^{64}$Ge predicts that the $1p1n$ alignment
takes place at $12_1^+$ just above the $gs$ band from $0_1^+$ to $10_1^+$
and continues till $18_1^+$, but there is no $2n$ aligned state
in the even-$J$ positive-parity yrast states.  For $^{70}$Ge, on the other
hand, the calculation yields only one $1p1n$ aligned yrast state, $16_1^+$.
The results suggest that the $1p1n$ aligned state is favored when the neutron
excess $N-Z$ is small, especially when $N=Z$.  This is probably related to
the existence of suitable low-energy states in the $A-2$ subsystem excluding
the $1p1n$ pair $(g_{9/2})^2_{J=9,T=0}$, for instance, such as the $T=0$
states of $^{62}$Ga for $^{64}$Ge and the $T=1$ states of $^{64}$Ga
for $^{66}$Ge.  We got the $T=0$ $1p1n$ aligned states $14_1^+$ and $16_1^+$
above the $gs$ band ($0_1^+ - 10_1^+$) for $^{62}$Zn, while we have no
$1p1n$ aligned state in the even-$J$ positive-parity yrast states for the
Zn isotopes with $A>62$.  It is interesting that the states $10_1^+$ and
$12_1^+$ are the $T=1$ $1p1n$ aligned states but the $14_1^+$ state is
the $T=0$ $1p1n$ aligned state in our calculation for $^{60}$Zn,
although the extended $P+QQ$ model may not be good enough for the four
valence-nucleon system \cite{Hase2}. The $^{68}$Se nucleus has the maximum
dimension in the present shell model calculations, i.e., about 
$1.6 \times 10^8$ for $0_1^+$.  The calculation predicts the early $T=0$
$1p1n$ alignment at $10_1^+$, namely, the $T=0$ $1p1n$ aligned states
from $10_1^+$ to $18_1^+$ above the $gs$ band ($0_1^+ - 8_1^+$).
We can expect the existence of the $T=0$ $1p1n$ aligned states in $^{70}$Se
and $^{72}$Se similar to $^{66}$Ge and $^{68}$Ge, but we did not make
calculations because of large dimensions.

   The expectation values of proton and neutron numbers in the four orbits
are tabulated also for the second band on $2_2^+$ observed in $^{66}$Ge,
in Table \ref{table3}.  Table \ref{table3} shows that the states
$2_2^+$, $4_2^+$, and $6_2^+$ (see Figs. \ref{fig1} and \ref{fig4}(a))
have no aligned nucleons in the $g_{9/2}$ orbit.
There is not a notable difference between the low-spin states up to $J=6$
of the $gs$ and second bands with respect to the distribution of nucleons.
The $8_2^+$ state has the aligned $2n$ pair in the $g_{9/2}$ orbit
as mentioned above.  The experiment for $^{68}$Ge \cite{Ward} found a cascade
of decay $8_2^+$ (or $8_3^+$)
 $\rightarrow 6_2^+ \rightarrow 4_3^+ \rightarrow 2_3^+ \rightarrow 0_2^+$
in addition to the sequence $6_2^+ \rightarrow 4_2^+ \rightarrow 2_2^+$.
A similar decay scheme could be found for $^{66}$Ge.
The low-spin excited band near above the $gs$ band should be investigated
further, but we leave the investigation except for the nuclear shape
discussed later.

\subsection{Particle alignments in other yrast states}

%===============  fig. 11  ========================================
\begin{figure}[b]
\includegraphics[width=7.0cm,height=12.0cm]{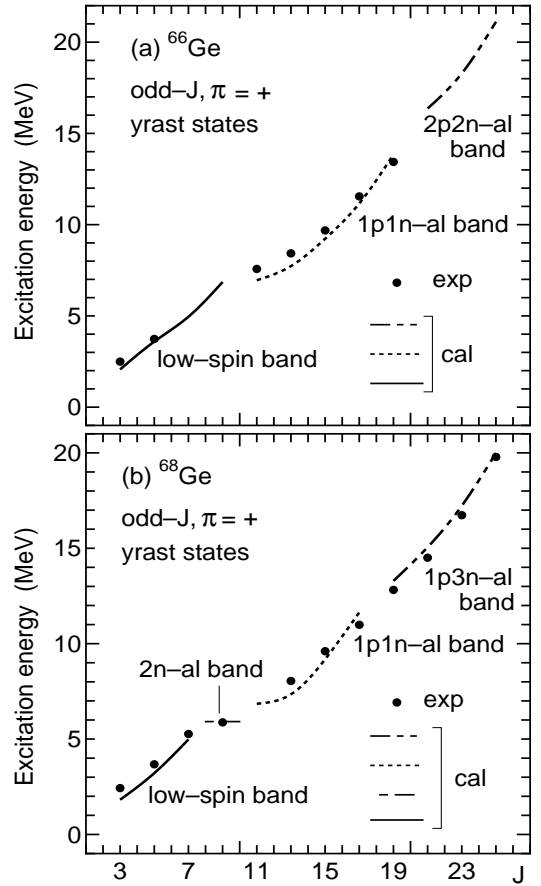}
  \caption{Comparison of the calculated odd-$J$ positive-parity yrast states
           (curves) with the experimental ones (solid circles)
           for (a) $^{66}$Ge and (B) $^{68}$Ge.}
  \label{fig11}
\end{figure}
%=================================================================

   Let us get onto the subject of other yrast states with positive and
negative parities ($\pi =\pm$) of $^{66}$Ge and $^{68}$Ge.

   The odd-$J$ positive-parity yrast states of $^{66}$Ge are interesting
because of the cascade $\Delta J=1$ transitions on the $11_1^+$ state
 \cite{Stefanova}.  We have discussed it in section~\ref{sec3.2}.
For these odd-$J$ states, we show calculated data about their wave functions
in Appendix A: the occupation probabilities and $Q$ moments in Table
\ref{tableAp1}; The expectation values of the spin and isospin of nucleons
occupying the $g_{9/2}$ orbit, $J_{g9/2}$ and $T_{g9/2}$, in Fig. \ref{fig17}.
Table \ref{tableAp1} shows that a structural change takes place at $11_1^+$
and the states $11_1^+$, $13_1^+$, $15_1^+$, $17_1^+$, and $19_1^+$ have
a common structure.  The common structure is created by the $T=0$, $J=9$
$1p1n$ alignment in the $g_{9/2}$ orbit, as known from Fig. \ref{fig17}.
We can now understand the almost equivalent two $J- \omega$ graphs
from $13_1^+$ to $19_1^+$ and from $14_1^+$ to $18_1^+$ in Fig. \ref{fig4}(a)
(see the dotted and solid lines).  The odd-$J$ states $11_1^+$ to $19_1^+$
have the structure of the $T=0$, $J=9$ $1p1n$ alignment which is the same
as that of the even-$J$ states $14_1^+$ to $18_1^+$.
It should be remembered that the $12_1^+$ state resembles the $1p1n$ aligned
state rather than the $2n$ aligned state.
  The cascade $\Delta J=1$ transitions
 $ 15_1^+ \rightarrow 14_1^+ \rightarrow 13_1^+ \rightarrow 
 12_1^+ \rightarrow 11_1^+ $
can be related to the common structure between the odd-$J$ and even-$J$
states in our model, which is in disagreement with the consideration
of possible four-quasiparticle structure in Ref. \cite{Stefanova}.
 The large value of $B(M1:12_1^+ \rightarrow 11_1^+)$ in our calculation
seems to be related to the strong mixing of the $1p1n$ and $2n$ aligned states
at $12_1^+$.  The mixing of the $1p1n$ and $2n$  aligned bands
could affect $B(M1)$ values for other states and hence the cascade
$\Delta J=1$ transitions.

   It is notable that the $T=1$, $J=8$ $2n$ aligned states do not become
the yrast states when $J={\rm odd}$ and $\pi =+$ in $^{66}$Ge.
This may be related to kinematic effects in the spin and isospin couplings.
The calculation predicts that the yrast states $21_1^+$, $23_1^+$, and $25_1^+$
are the $2p2n$ aligned states with $J_{g9/2} \approx 16$ and $T_{g9/2} \approx 0$.
 We have the $1p1n$ aligned band on $11_1^+$ and the $2p2n$ aligned band
on $21_1^+$ in addition to the low-spin band with no alignment,
 for the odd-$J$ positive-parity yrast states of $^{66}$Ge.
Let us take a glance at the $E_x-J$ graph in Fig. \ref{fig11}(a) to compare it
with a corresponding figure of $^{68}$Ge in the lower panel (b). 
Although the experimental data are not sufficient for comparison,
the good reproduction of the $J- \omega$ graph from $13_1^+$ to $17_1^+$
in Fig. \ref{fig4}(a) supports our classification.

   For $^{68}$Ge, odd-$J$ positive-parity yrast states are experimentally
observed up to $25_1^+$.  We show calculated data about their wave functions
in the same manner as that for $^{66}$Ge, in Table \ref{tableAp2} and
Fig. \ref{fig18} of Appendix A.  Table \ref{tableAp2} indicates successive
three types of particle aligned bands above the low-spin band with no alignment.
They are the $T=1$, $J=8$ $2n$ aligned band which has only one yrast state
$9_1^+$, the $T=0$, $J=9$ $1p1n$ aligned band from $11_1^+$ to $17_1^+$
and the one-proton-three-neutron ($1p3n$) aligned band
from $19_1^+$ to $25_1^+$, as known from Fig. \ref{fig18}.
  Interestingly, the two excess neutrons in $^{68}$Ge as compared with $^{66}$Ge
produce another type of particle alignment in the $g_{9/2}$ orbit
for the odd-$J$ positive-parity yrast states with $J\ge 19$.
The $1p3n$ aligned states have the combination of $T=1$ $2n$ pair and
$T=0$ $pn$ pair in the $g_{9/2}$ orbit which produces the spin
 $J_{g9/2}=9/2^\pi +(9/2+7/2+5/2)^\nu =15$ and the isospin $T_{g9/2}=1$.
The calculation reproduces well the experimental $J-\omega$ graph for
the $J\ge 15$ states except $17_1^+$, as mentioned for Fig. \ref{fig5}(a).
Thus, we have the four bands for the odd-$J$ positive-parity yrast states of
$^{68}$Ge, as shown in the $E_x-J$ graph of Fig. \ref{fig11}(b).
The calculation traces quite well the experimental footprints
in Fig. \ref{fig11}(b).

   There are sufficient data on the negative-parity yrast states of $^{66}$Ge
and $^{68}$Ge except for $J={\rm even}$ of $^{66}$Ge.  The negative-parity
states have odd-number nucleons in the $pf$ shell with $\pi =-$ and hence
at least one nucleon must occupy the $g_{9/2}$ orbit in the even-mass
Ge isotopes.  This condition produces different structures
from those of the positive-parity states in $^{66}$Ge and $^{68}$Ge.

%===============  fig. 12  ========================================
\begin{figure}[b]
\includegraphics[width=7.0cm,height=12.0cm]{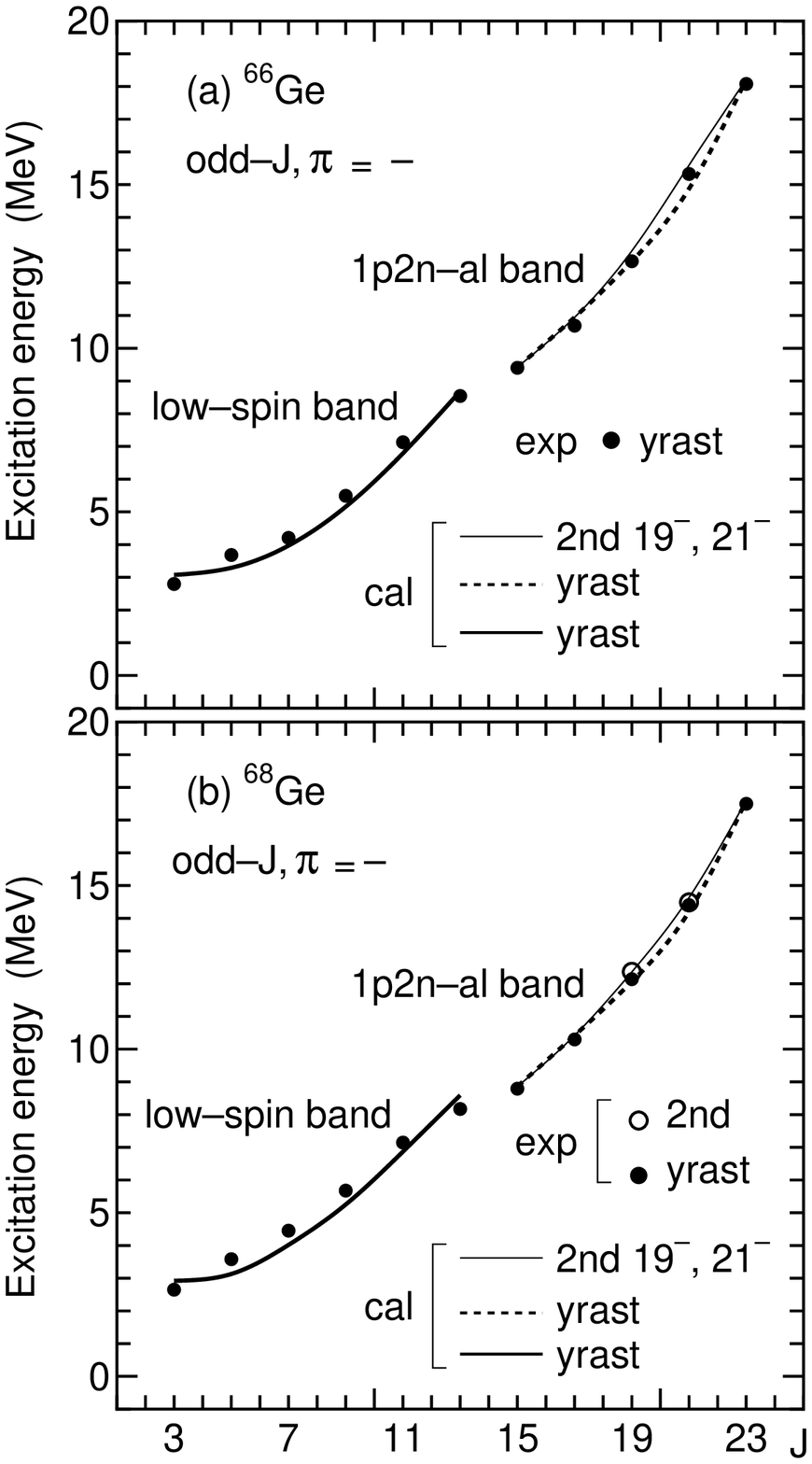}
  \caption{Comparison of the calculated odd-$J$ negative-parity states
           (curves) with the experimental ones (circles)
           for (a) $^{66}$Ge and (b) $^{68}$Ge.}
  \label{fig12}
\end{figure}
%=================================================================

   First, we consider the odd-$J$ negative-parity yrast states of $^{66}$Ge
which are experimentally observed up to tentative $23_1^-$.
We show calculated data about their wave functions in Table \ref{tableAp3}
and Fig. \ref{fig19} of Appendix B.  Table \ref{tableAp3} indicates
that the low-spin states up to $13_1^-$ have approximately one nucleon
in the $g_{9/2}$ orbit, giving the spin $J_{g9/2} \approx 9/2$ and
the isospin $T_{g9/2} \approx 1/2$.
At $15_1^-$, one proton and two neutrons ($1p2n$) align in the $g_{9/2}$ orbit
and produce the spin $J_{g9/2}=9/2^\pi + (9/2+7/2)^\nu =25/2$ and
the isospin $T_{g9/2}=1/2$.
The $1p2n$ aligned band continues up to $23_1^-$ where the band terminates.
The $1p2n$ aligned states can be regarded as the $T=1$ aligned $2n$ pair
coupled with one proton and also as the $T=0$ aligned $pn$ pair coupled with
one neutron. We cannot distinguish the two types of coupling.
By the way, the $2p2n$ aligned states are forbidden by the condition
that the nucleon number in the $g_{9/2}$ orbit must be odd.
The next particle-aligned state is the two-proton-three-neutron ($2p3n$) aligned
state $25_1^-$ as predicted in the bottom line of Table \ref{tableAp3}.
The sequence of the odd-$J$ negative-parity yrast states has two bands,
the low-spin band with one neutron in the $g_{9/2}$ orbit and the $1p2n$
aligned band on $15_1^-$, which is illustrated in Fig. \ref{fig12}(a).
The theoretical two bands nicely trace the experimental footprints.
  We have discussed in section~\ref{sec3.2} that the calculated $B(E2)$ values
recommend us to regard the $19_2^-$ and $21_2^-$ states as the members
of the band on $15_1^-$ (see Fig. \ref{fig1}).  We plot two curves for the band
on $15_1^-$ in Fig. \ref{fig12}(a). One is the curve connecting the yrast states
$15_1^-$ to $23_1^-$ and the other is the curve passing the states $19_2^-$
and $21_2^-$.

   Secondly, we consider the odd-$J$ negative-parity yrast states of $^{68}$Ge.
The calculated data about their wave functions  are given in Table \ref{tableAp4}
and Fig. \ref{fig20} of Appendix B.  Table \ref{tableAp4} and Fig. \ref{fig20}
show that the odd-$J$ negative-parity yrast states of $^{68}$Ge have
essentially the same features as those of $^{66}$Ge with respect to the
particle alignments in the $g_{9/2}$ orbit.  There are two bands, the low-spin
band with one neutron in the $g_{9/2}$ orbit and the $1p2n$ aligned band
 which starts from $15_1^-$ and terminates at $23_1^-$.
In the $1p2n$ aligned band, the three nucleons in the $g_{9/2}$ orbit have
$J_{g9/2} \approx 25/2$ and $T_{g9/2} \approx 1/2$.
 We show the $E_x-J$ graph in Fig. \ref{fig12}(b).  The agreement between theory
and experiment is good especially for the $1p2n$ aligned band on $15_1^-$.  
We have discussed the continuation of bands near $19_1^-$ and $21_1^-$
in section~\ref{sec3.2}.  The calculated $B(E2)$ values have recommended us
to regard the states $17_1^-$, $19_2^-$, $21_2^-$, and $23_1^-$
as members of the band on $15_1^-$. The curve connecting these states appears
to be smoother in Fig. \ref{fig12}(b), although this assignment is not necessarily
consistent with the assignment of bands in the experiment \cite{Ward}.
We note that the $19_2^-$ and $21_2^-$ states have the $1p2n$ alignment
in the $g_{9/2}$ orbit.

%===============  fig. 13  ========================================
\begin{figure}[t]
\includegraphics[width=7.0cm,height=7.0cm]{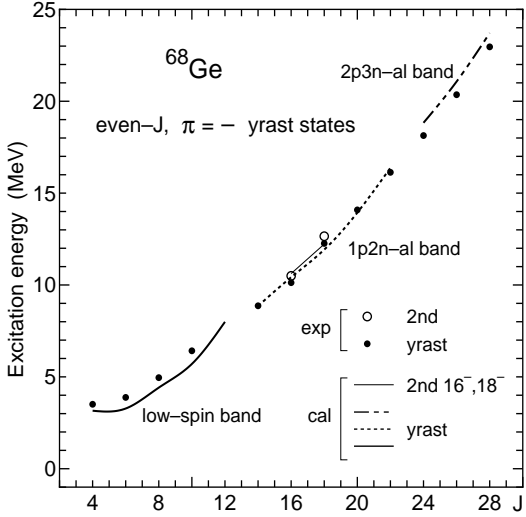}
  \caption{Comparison of the calculated even-$J$ negative-parity states
           (curves) with the experimental ones (circles) for $^{68}$Ge.}
  \label{fig13}
\end{figure}
%=================================================================

  Thirdly, let us look at the even-$J$ negative-parity yrast states of $^{68}$Ge.
The maximum spin $J^\pi =28^-$ is experimentally observed in this sequence.
We show calculated data about their wave functions in Table \ref{tableAp5}
and Fig. \ref{fig21} of Appendix B.  Table \ref{tableAp5} and Fig. \ref{fig21}
indicate that the even-$J$ negative-parity yrast states up to $22_1^-$
have a similar nature to the odd-$J$ negative-parity yrast states
up to $23_1^-$ in $^{68}$Ge.  The calculated high-spin states from $14_1^-$
to $22_1^-$ contain the aligned three nucleons ($1p2n$)
with $J_{g9/2} \approx 25/2$ and $T_{g9/2} \approx 1/2$ in the  $g_{9/2}$ orbit.
This is consistent with the discussion in Ref. \cite{Ward} that there are
very low-lying aligned $14^-$ and $16^-$ states with one $g_{9/2}$ proton and
two $g_{9/2}$ neutrons.
The calculated $B(E2)$ values for the $18^- \rightarrow 16^-$ transitions are
0.7 $e^2$fm$^4$ for $18_1^- \rightarrow 16_1^-$ and
 209 $e^2$fm$^4$ for $18_1^- \rightarrow 16_2^-$, which suggests discontinuity
 of the $1p2n$ aligned states between $16_1^-$ and $18_1^-$.
This is also consistent with no detection of the $18_1^- \rightarrow 16_1^-$ 
transition in the experiment \cite{Ward}.
For the higher-spin states $24_1^-$, $26_1^-$, and $28_1^-$,
 Fig. \ref{fig21} indicates the $2p3n$ alignment coupled to 
$J_{g9/2}=(9/2+7/2)^\pi +(9/2+7/2+5/2)^\nu =37/2$ and $T_{g9/2}=1/2$
in the $g_{9/2}$ orbit.
These states are the members of the $2p3n$ aligned band.
This assignment is supported by the calculated $B(E2)$ values,
a small value for the transition $24_1^- \rightarrow 22_1^-$ (which shows
discontinuity between the states $22_1^-$ and $24_1^-$) and large values
for the transitions $28_1^- \rightarrow 26_1^- \rightarrow 24_1^-$.
 Tables \ref{tableAp3} and \ref{tableAp4} show that there could be
the $2p3n$ aligned states with $J={\rm odd}$, $\pi =-$ above
$23_1^-$ both in $^{66}$Ge and $^{68}$Ge.
We have three bands in the even-$J$ negative-parity yrast states of $^{68}$Ge,
the low-spin band with one neutron in the $g_{9/2}$ orbit,
the $1p2n$ aligned band on $14_1^-$ and the $2p3n$ aligned band on $24_1^-$,
as shown in the $E_x-J$ graph of Fig. \ref{fig13}.
 The agreement between theory and experiment
is not good for the low-spin band but is satisfactorily good for the $1p2n$
and $2p3n$ aligned bands.

   We close this subsection by noting that the structural changes at the band
crossings in Figs. \ref{fig11}-\ref{fig13} manifest themselves in the changes
of the spectroscopic $Q$ moment (see Tables \ref{tableAp1}-\ref{tableAp5}
in Appendixes A and B).

%=================================================================
\section{Dependence of the first band crossing on neuron number}\label{sec5}

%===============  fig. 14  ========================================
\begin{figure}[b]
\includegraphics[width=8.0cm,height=12.0cm]{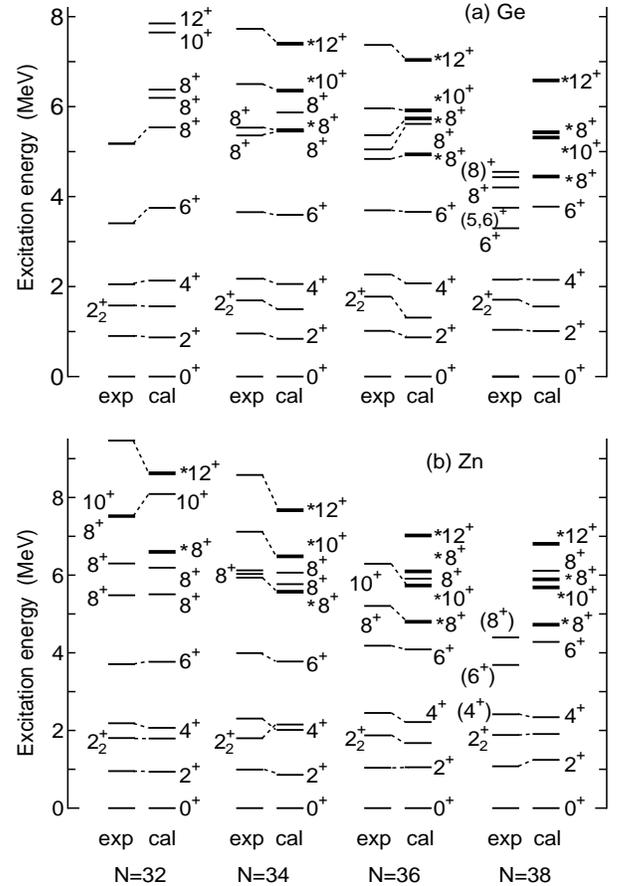}
  \caption{Changes of the low-energy levels in the (a) Ge and (b) Zn isotopes.
           The $2n$ and $1p1n$ aligned states (thick lines) are distinguished
           with the asterisks from the members of the $gs$ band.}
  \label{fig14}
\end{figure}
%=================================================================

   We have seen the change of the first band crossing (backbending point) 
from $^{66}$Ge to $^{68}$Ge.  The backbending takes place at $10_1^+$
in $^{66}$Ge and $8_1^+$ in $^{68}$Ge.  This phenomenon is explained in terms
of the $2n$ alignment in the previous section.  Namely, it is caused
by the competition between the $gs$ band and the $2n$ aligned band
for holding the lowest (yrast) position in energy near $J=8$.
 Let us illustrate the competition in the Ge isotopes including $^{64}$Ge
 and $^{70}$Ge, in Fig. \ref{fig14}(a).  We calculated energy levels
of $^{64}$Ge and $^{70}$Ge using the same Hamiltonian as that for $^{66}$Ge
and $^{68}$Ge.  The force strengths in Eq. (\ref{eq:3}) are too strong
for the $gs$ band of the $N=Z$ nucleus $^{64}$Ge, pushing up the energy
levels as $J$ increases.  The present model is, however, good enough to discuss
the band crossing near $J=8$.  The $2n$ aligned states $8^+$ and $10^+$,
and the $12^+$ state with dominant component of the $1p1n$ alignment
are denoted by the thick lines with the asterisks (as $*8^+$)
in Fig. \ref{fig14}(a).

   In Fig. \ref{fig14}(a), while the experimental $6_1^+$ state goes up gradually
till $^{68}$Ge with $N=36$, the $8_1^+$ state reaches the peak at $^{66}$Ge
with $N=34$ and goes down as $N$ increases.  (The $6_1^+$ state is not
successfully reproduced for $^{70}$Ge with $N=38$.)
 The calculation indicates that the $2n$ aligned $8_1^+$ state does not
intrude among the lowest three $8^+$ states when $N=32$, but competes with
the $8^+$ member of the $gs$ band when $N=34$ and becomes lower in energy
than the latter when $N \ge 36$.  
It should be noticed that the $N$-dependent behaviors of energy levels
of the Ge isotopes are reproduced by the $N$-independent parameters
of interactions.
We can say that the first band crossing depends on the neutron number.
This feature can be attributed to the upward movement of the neutron Fermi
level with the increase in the neutron number $N$.
The approach of the Fermi level to the $g_{9/2}$ orbit makes nucleons be easy
to go into the $g_{9/2}$ orbit and to get high spin by particle alignment.

   The same behavior is observed in the Zn isotopes as shown
in Fig. \ref{fig14}(b).  Here, we strengthened the force strengths
for the Zn isotopes by setting $A=62$ in Eq. (\ref{eq:3}).
We use the same ($N$-independent) force strengths for the Zn isotopes
with $A=62 - 68$.
The experimental $6_1^+$ state goes up gradually till $N=36$.  The $8_1^+$
 state reaches the peak at $N=34$ and goes down as $N$ increases. 
The change in the Zn isotopes shows a slight difference from that
in the Ge isotopes.  In $^{64}$Zn with $N=34$, the $2n$ aligned $8^+$ state
is lower than the $8^+$ member of the $gs$ band, but the backbending
takes place at $10_1^+$ like $^{66}$Ge with $N=34$.
The backbending at $8_1^+$ is observed at $N=36$ also in the Zn isotopes.
 We can see the same behavior of the $8_1^+$ state in the Se isotopes
as those in the Zn and Ge isotopes.
The explanation for the backebnding at $8_1^+$ when $N=36$ mentioned
for the Ge isotopes, that the position of the neutron Fermi level affects
whether the $2n$ alignment in the $g_{9/2}$ orbit is easy or not,
is therefore reasonable.  
The common behavior depending on the neutron number supports our conclusion
that the first band crossing (backbending) at $8_1$ is caused by the $2n$
alignment in the $g_{9/2}$ orbit.

%=================================================================
\section{Nuclear shapes of Ge isotopes}\label{sec6}

   The nuclear shape or the coexistence of oblate and prolate shapes
has been a hot topic in nuclei around Ge isotopes
 \cite{Stefanova,Bengtsson,Nazarewicz,Sarri,Ennis,Yamagami}.
The VAMPIR calculations gave positive (oblate) $Q$ moments to the $gs$
band and negative (prolate) $Q$ moments to the second low-spin band 
in $^{68}$Ge \cite{Petrovici1,Petrovici2,Petrovici3} . 
The recent calculations \cite{Stefanova,Sarri} also yielded
the deepest minimum for an oblate shape and the next minima for prolate shapes
in $^{68}$Ge and $^{66}$Ge, and predicted shape coexistence and $\gamma$
softness.

   In order to investigate this problem, we calculated spectroscopic $Q$
moments of all the states of $^{66}$Ge and $^{68}$Ge, which are listed
in Tables \ref{table3}-\ref{tableAp5}.
Our model yields negative $Q$ moments for the $gs$ band and other states,
except that the $2_2^+$ and $4_3^+$ states of $^{66}$Ge and the $2_2^+$,
$4_3^+$, and $6_2^+$ states of $^{68}$Ge have positive $Q$ moments. 
 In this sense, we get different shapes with opposite signs of $Q$ moments
in the low-lying states of $^{66}$Ge and $^{68}$Ge, but the positive $Q$
moment narrowly appears in the second or third excited state
($2^+$, $4^+$, or $6^+$) in our calculations.
This suggests that the oblate minimum is shallower than the prolate minimum
in the potential energy surface.  Our result disagrees with the VAMPIR
result \cite{Petrovici1,Petrovici2,Petrovici3}.
A notable difference between our model and the VAMPIR model is
in the single-particle energies.  The $g_{9/2}$ level lies below the $p_{1/2}$
level in the VAMPIR model.  We calculated the spectroscopic $Q$ moment by setting
$\varepsilon_{g9/2}=\varepsilon_{p1/2}$, but could not get positive $Q$ moments
for the $gs$ band.  We also examined the $Q$ moment by changing the force
parameters.  However, changing the parameters does not affect
the signs of the $Q$ moments within the limit that the energy
levels are not seriously damaged.

%===============  fig. 15  ========================================
\begin{figure}[b]
\includegraphics[width=6.5cm,height=9.2cm]{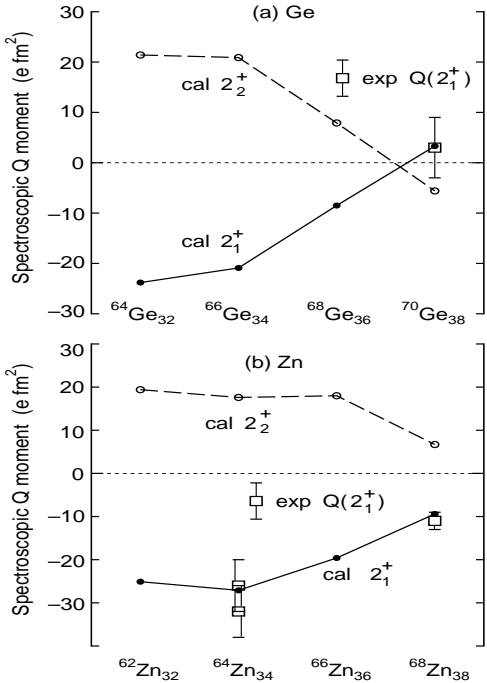}
  \caption{Comparison between calculated and experimental $Q$ moments
           in the (a) Ge and (b) Zn isotopes.}
  \label{fig15}
\end{figure}
%=================================================================

   The $Q$ moment of the $2_1^+$ state ($Q(2_1^+)$) has not experimentally
been observed for $^{66}$Ge and $^{68}$Ge yet, but was measured for $^{70}$Ge.
To examine our model and to see the changes of $Q(2_1^+)$ and $Q(2_2^+)$
depending on the neutron number, we calculated $Q$ moments for $^{70}$Ge and
$^{64}$Ge.  The calculation for $^{64}$Ge yields negative $Q$ moments
for the $gs$ band and exceptionally yields a positive $Q$ moment
for the $2_2^+$ state, which is similar to the results for $^{66}$Ge and
$^{68}$Ge.  In $^{70}$Ge, on the contrary, the $2_1^+$ and $4_1^+$ states
have positive $Q$ moments, while the $2_2^+$ state has a negative $Q$ moment.
Let us illustrate the changes of $Q(2_1^+)$ and $Q(2_2^+)$ in the Ge isotopes
from $^{64}$Ge ($N=32$) to $^{70}$Ge ($N=38$), in Fig. \ref{fig15}(a).
The $Q(2_1^+)$ value calculated for $^{70}$Ge is in good agreement with
the experimental value, in Fig. \ref{fig15}(a).
This agreement suggests that the present predictions for $Q(2_1^+)$ and
$Q(2_2^+)$ are better than those of Refs. \cite{Petrovici1,Petrovici2,Petrovici3}.
There are two data on $Q(2_1^+)$ for Zn isotopes, {\it i.e.,} for $^{64}$Zn
and $^{68}$Zn.
We calculated $Q$ moments for the Zn isotopes from $^{62}$Zn ($N=32$)
to $^{68}$Zn ($N=38$).  Calculated results are shown in Fig. \ref{fig15}(b).
This figure also indicates the success of our model in reproducing
the experimental $Q(2_1^+)$ values of $^{64}$Zn and $^{68}$Zn, which supports
our prediction for the $Q$ moments of the Ge isotopes.
 It is probable that in contrast to the suggestions of Refs.
 \cite{Stefanova,Petrovici1,Petrovici2,Petrovici3}
 the $2_1^+$ state of the $gs$ band has a negative $Q$ moment,
 while the $2_2^+$ state has a positive $Q$ moment, in $^{66}$Ge and $^{68}$Ge.

   Figure \ref{fig15}(a) suggests a gradual decrease of the prolate (oblate)
deformation of the $2_1^+$ ($2_2^+$) state from $^{66}$Ge to $^{68}$Ge, and
a kind of shape inversions of the $2_1^+$ and $2_2^+$ states in $^{70}$Ge.
Similar decreases of the prolate and oblate deformations of the $2_1^+$
and $2_2^+$ states from $^{64}$Zn to $^{68}$Zn are seen in Fig. \ref{fig15}(b).
There seems to be a trend that the prolate (oblate) shape of the $2_1^+$
($2_2^+$) state tends toward the oblate (prolate) shape when the neutron
number goes over $N=34$.  This trend is clearer in the Ge isotopes
with two more protons than in the Zn isotopes.  It, therefore, must be
related also to the proton number.
In Ref. \cite{Kaneko2}, we investigated the shape transition
from the prolate shape of $^{64}$Ge to the oblate shape of $^{68}$Se,
and showed that the increase of the nucleon number in the $2p_{1/2}$ orbit
($\langle n_{p1/2} \rangle $) caused by the monopole correction
$H^{T=1}_{\rm mc}(f_{5/2},p_{1/2})$ contributes to the oblate deformation.
The acceleration of the shape change from the Zn isotopes to the Ge
isotopes can be attributed to the occupation of the $2p_{1/2}$ orbit
by protons as well as neutrons.  The shape change seen in Fig. \ref{fig15}(a)
(Fig. \ref{fig15}(b)) corresponds to the increase of the neutron number
in the $2p_{1/2}$ orbit, as shown in Tables \ref{table3}
and \ref{table4} (incidentally, $\langle n_{p1/2}^\nu \rangle$ is 1.48
for the $2_1^+$ state of $^{70}$Ge).
This must be related to the upward movement of the neutron Fermi level
from $N=34$ to $N=38$, which causes the lowering of the $2n$ aligned $8_1^+$
state from $^{66}$Ge to $^{70}$Ge as discussed in the last section.

%===============  fig. 16  ========================================
\begin{figure}[b]
\includegraphics[width=6.5cm,height=10.0cm]{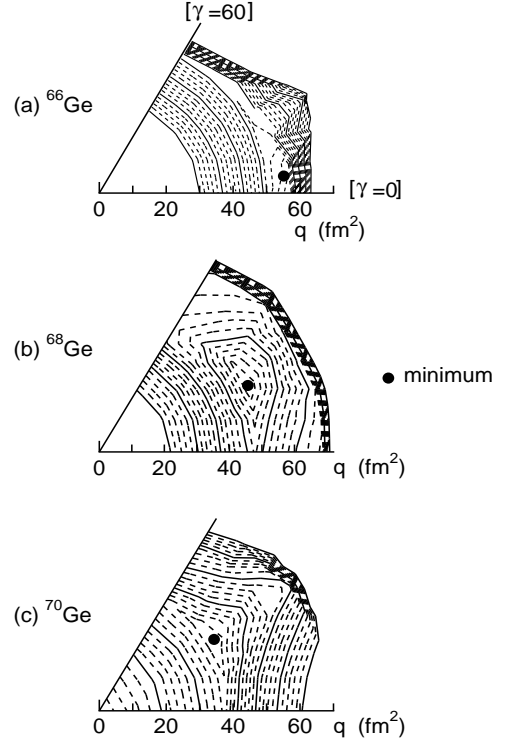}
  \caption{The energy surface
           $ \langle q, \gamma | H | q, \gamma \rangle _{J=0} $
           in the $q-\gamma$ plane ($0^\circ \le \gamma \le 60^\circ$)
           plotted with contours,
           for (a) $^{66}$Ge, (b) $^{68}$Ge, and (c) $^{70}$Ge.}
  \label{fig16}
\end{figure}
%=================================================================

   Let us investigate further the nuclear shapes of the Ge isotopes and
the triaxiality by using an alternative method, the CHF method
\cite{Mizusaki2,Hara}.  We consider the constrained Hamiltonian with
the following quadratic terms of quadrupole moment $q_M$ and spin $J$:
\begin{eqnarray}
 & {} & H ^\prime =  H
       + c_1 \sum_{M=0,\pm 2} ( \langle Q_{2M} \rangle - q_M )^2  \nonumber \\
 & {} & \hspace{1.3cm} + c_2 [ \langle J_x \rangle - \sqrt{J(J+1)} ]^2 ,
         \label{eq:5} \\
 & {} & q_0 = \sqrt{ {5 \over 4 \pi} } q \cos \gamma , \quad
        q_{\pm 2} = \sqrt{ {5 \over 8 \pi} } q \sin \gamma ,        \label{eq:6}
\end{eqnarray}
where $c_1$ and $c_2$ are predefined positive constants, and $J_x$ denotes
the $x$-component of the angular momentum operator.
We plot the energy surface $ \langle q, \gamma | H | q, \gamma \rangle _{J=0}$
in the $q-\gamma$ plane for $^{66}$Ge, $^{68}$Ge, and $^{70}$Ge,
in Fig. \ref{fig16}.  In Fig. \ref{fig16}(a) for $^{66}$Ge, the energy surface
has a minimum near $q \approx 56$ fm$^2$ and $\gamma \approx 0^\circ$, 
suggesting an axially symmetric prolate deformation with $\beta \approx 0.2$. 
This value $\beta \approx 0.2$, which is consistent with the calculated $B(E2)$
value $B(E2:2_1^+ \rightarrow 0_1^+) \approx 281$ $e^2$ fm$^4$
on the assumption of the axial symmetry, corresponds with
previous predictions $\beta \approx 0.2 - 0.22$ \cite{Stefanova,Ennis}.
The energy surface in Fig. \ref{fig16}(a), however, displays a valley
along the $\gamma$ direction, which means $\gamma$ softness of $^{66}$Ge
like $^{64}$Ge (see Ref. \cite{Kaneko2}).
The energy surface for $^{68}$Ge indicates a triaxial deformation 
in Fig. \ref{fig16}(b).  The minimum is near $q \approx 50$ fm$^2$ and
$\gamma \lesssim 30^\circ$ in the $q-\gamma$ plane. This seems to be consistent
with the small negative value of $Q(2_1^+)$ obtained by the shell model
calculation, from the relation $Q \propto -q \cos{(3 \gamma)}$
in the Davydov model.
The energy surface for $^{70}$Ge also shows a triaxial deformation
in Fig. \ref{fig16}(c).  The minimum near $q \approx 45$ fm$^2$ and
$\gamma \gtrsim 30^\circ$ in the $q-\gamma$ plane corresponds to the small
positive value of the experimental $Q$ moment $Q(2_1^+)$.

  The CHF calculations thus support the above discussions which are based on
the shell model calculations of $Q(2_1^+)$.
The energy surface calculations interpret the shape change depending on
the neutron number from $^{66}$Ge to $^{70}$Ge as the movement along the $\gamma$
direction (crossing the $\gamma =30^\circ$ border) with a gradual decrease
of quadrupole deformation in the $q-\gamma$ plane.
A similar trend is seen in the CHF results for the Zn isotopes from $^{64}$Zn
to $^{68}$Zn, although the minimum point for $^{68}$Zn does not cross
the $\gamma =30^\circ$ border.
Our figures of the energy surfaces for $^{66}$Ge and $^{68}$Ge do not correspond
with those of the TRS calculations \cite{Stefanova}.  The discrepancy is
probably due to the difference between the two Hamiltonians used
in our calculations and the TRS ones.
Our Hamiltonian describes well the energy levels and other properties
of $^{66}$Ge and $^{68}$Ge, and also the $Q$ moments observed in $^{70}$Ge,
$^{64}$Zn, and $^{68}$Zn, as we have seen.

%=================================================================
\section{Summary}\label{sec7}

   We have investigated the structure of $^{66}$Ge and $^{68}$Ge,
by carrying out large scale shell model calculations with the extended
$P+QQ$ Hamiltonian in the configuration space
($2p_{3/2}$, $1f_{5/2}$, $2p_{1/2}$, $1g_{9/2}$).
The shell model reproduces excellently the energy levels of $^{66}$Ge and
$^{68}$Ge, and also satisfactorily those of $^{64}$Ge, $^{65}$Ge, $^{67}$Ge,
and $^{70}$Ge ($^{62}$Zn, $^{64}$Zn, $^{66}$Zn, and $^{68}$Zn). 
The model explains well the graphs of spin versus angular frequency for the
positive- and negative-parity yrast states with even $J$ and odd $J$,
and quite well the experimental $B(E2)$ values, in $^{66}$Ge and $^{68}$Ge.
The calculated $B(E2)$ values are basically consistent with the band schemes
assigned by $\gamma$ transitions in the experiments \cite{Stefanova,Ward}.

   To analyze the structure, we calculated the expectation values of proton
and neutron numbers in the four orbits, the expectation values of the spin
and isospin of nucleons in the $pf$ shell 
($2p_{3/2}$, $1f_{5/2}$, $2p_{1/2}$) and in the $g_{9/2}$ orbit, and the
spectroscopic $Q$ moment.  The analysis has clarified that the structural
changes in the four sequences of the positive- and negative-parity yrast states
with even $J$ and odd $J$ are caused by various types of particle alignments
in the $g_{9/2}$ orbit.
Although not all the predicted aligned states have been found, 
the calculations explain the experimental graphs of the excitation
energy versus spin as shown in Figs. \ref{fig6}, \ref{fig9},
 \ref{fig11}-\ref{fig13}.
The results are summarized as follows. \\
(A) $^{66}$Ge: \\
 (1) The even-$J$, $\pi =+$ sequence has the four bands up to $26_1^+$:
  the $gs$ band, $2n$ aligned band, $1p1n$ aligned band, and $2p2n$ aligned band.
 (2) The odd-$J$, $\pi =+$ sequence has the three bands up to $25_1^+$:
  the low-spin band, $1p1n$ aligned band, and $2p2n$ aligned band.
 (3) The odd-$J$, $\pi =-$ sequence has the two bands up to $23_1^-$:
  the low-spin band and $1p2n$ aligned band. \\
(B) $^{68}$Ge: \\
 (1) The even-$J$, $\pi =+$ sequence has the four bands up to $26_1^+$:
  the $gs$ band, $2n$ aligned band, $1p1n$ aligned band, and $2p2n$ aligned band.
 (2) The odd-$J$, $\pi =+$ sequence has the four bands up to $25_1^+$:
  the low-spin band, $2n$ aligned state, $1p1n$ aligned band,
  and $1p3n$ aligned band.
 (3) The odd-$J$, $\pi =-$ sequence has the two bands up to $23_1^-$:
  the low-spin band and $1p2n$ aligned band.
 (4) The even-$J$, $\pi =-$ sequence has the three bands up to $28_1^-$:
  the low-spin band, $1p2n$ aligned band, and $2p3n$ aligned band.

   The backbending takes place at $10_1^+$ in $^{66}$Ge and at $8_1^+$ in
$^{68}$Ge, which is explained by crossing of the $2n$ aligned band and
the $gs$ band in our calculations.  We have discussed that the change of
the first band crossing depending on the neutron number, which is observed
in the Ge, Zn, and Se isotopes, can be attributed to the movement of
the neutron Fermi level toward the $g_{9/2}$ orbit.

   We have looked into the nuclear shapes of the Ge isotopes at low energy
where no particle alignment takes place. The calculations predict
that the first and second $2^+$ states have opposite signs of spectroscopic
$Q$ moments in the Ge isotopes as follows.
The $2_1^+$ state has a negative $Q$ moment in $^{66}$Ge and $^{68}$Ge
and has a positive $Q$ moment in $^{70}$Ge,
while the $2_2^+$ state has an opposite sign of the $Q$ moment
against the $2_1^+$ state in these nuclei.
The prolate (oblate) deformation of the $2_1^+$ ($2_2^+$) state decreases
from $^{66}$Ge to $^{68}$Ge.
The CHF calculations interpret the shape change depending on the neutron
number as the movement along the $\gamma$ direction
(crossing the $\gamma =30^\circ$ border)
with a gradual decrease of quadrupole deformation in the $q-\gamma$ plane.
It is predicted that the ground state of $^{66}$Ge is prolate
and those of $^{68}$Ge and $^{70}$Ge are triaxial.
The prediction is supported by the agreement between the calculated
and experimental $Q$ moments ($Q(2_1^+)$) in $^{70}$Ge, $^{64}$Zn,
and $^{68}$Zn.

   The $1f_{7/2}$ orbit is not included in our calculations
because the extension of configuration space makes the shell model calculations
impossible.  This truncation ought to have effects on the results obtained. 
The $2d_{5/2}$ orbit possibly affects the results at high energy.
Better interaction parameters would be necessary for describing nuclei
in a wider region.

\newpage

%=================================================================
\appendix

\section{Structure of the odd-$J$ positive-parity yrast states}

%===============  table Ap1  ========================
\begin{table}[h]
\caption{The proton and neutron numbers $\langle n_a^\pi \rangle$ and
         $\langle n_a^\nu \rangle$, and calculated $Q$ moments (in $e$ fm$^2$),
         for the odd-$J$ positive-parity yrast states of $^{66}$Ge.}
\begin{tabular}{c|cccc|cccc|c}   \hline
        & \multicolumn{4}{c}{proton} & \multicolumn{4}{|c|}{neutron} & \\ \hline
$^{66}$Ge & $p_{3/2}$ & $f_{5/2}$ & $p_{1/2}$ & $g_{9/2}$
         & $p_{3/2}$ & $f_{5/2}$ & $p_{1/2}$ & $g_{9/2}$ & $Q$  \\ \hline
  $3_1^+$  & 1.79 & 1.56 & 0.56 & 0.10 & 2.44 & 2.72 & 0.70 & 0.14 &  -0.9 \\
  $5_1^+$  & 1.71 & 1.79 & 0.43 & 0.07 & 2.35 & 2.83 & 0.69 & 0.12 & -17.9 \\
  $7_1^+$  & 1.75 & 1.76 & 0.41 & 0.09 & 2.21 & 2.87 & 0.78 & 0.13 &  -7.3 \\
  $9_1^+$  & 1.76 & 1.80 & 0.35 & 0.09 & 2.21 & 2.87 & 0.78 & 0.14 & -14.8 \\ \hline
 $11_1^+$  & 1.24 & 1.14 & 0.61 & 1.02 & 1.90 & 2.49 & 0.56 & 1.06 & -77.9 \\
 $13_1^+$  & 1.30 & 1.12 & 0.57 & 1.01 & 1.79 & 2.58 & 0.59 & 1.04 & -82.6 \\
 $15_1^+$  & 1.30 & 1.14 & 0.55 & 1.01 & 1.79 & 2.45 & 0.72 & 1.04 & -86.4 \\
 $17_1^+$  & 1.24 & 1.17 & 0.58 & 1.01 & 1.77 & 2.22 & 0.97 & 1.03 & -89.0 \\
 $19_1^+$  & 1.23 & 1.16 & 0.59 & 1.02 & 1.67 & 2.17 & 1.13 & 1.04 & -91.2 \\ \hline
 $21_1^+$  & 0.67 & 0.93 & 0.49 & 1.92 & 1.53 & 1.71 & 0.65 & 2.11 & -81.4 \\
 $23_1^+$  & 0.60 & 1.06 & 0.39 & 1.96 & 1.36 & 1.94 & 0.65 & 2.05 & -83.4 \\
 $25_1^+$  & 0.46 & 1.07 & 0.51 & 1.96 & 1.37 & 2.06 & 0.50 & 2.07 & -84.3 \\ \hline
\end{tabular}
\label{tableAp1}
\end{table}
%=========================================================

%===============  fig. 17  ========================================
\begin{figure}[h]
\includegraphics[width=6.0cm,height=6.2cm]{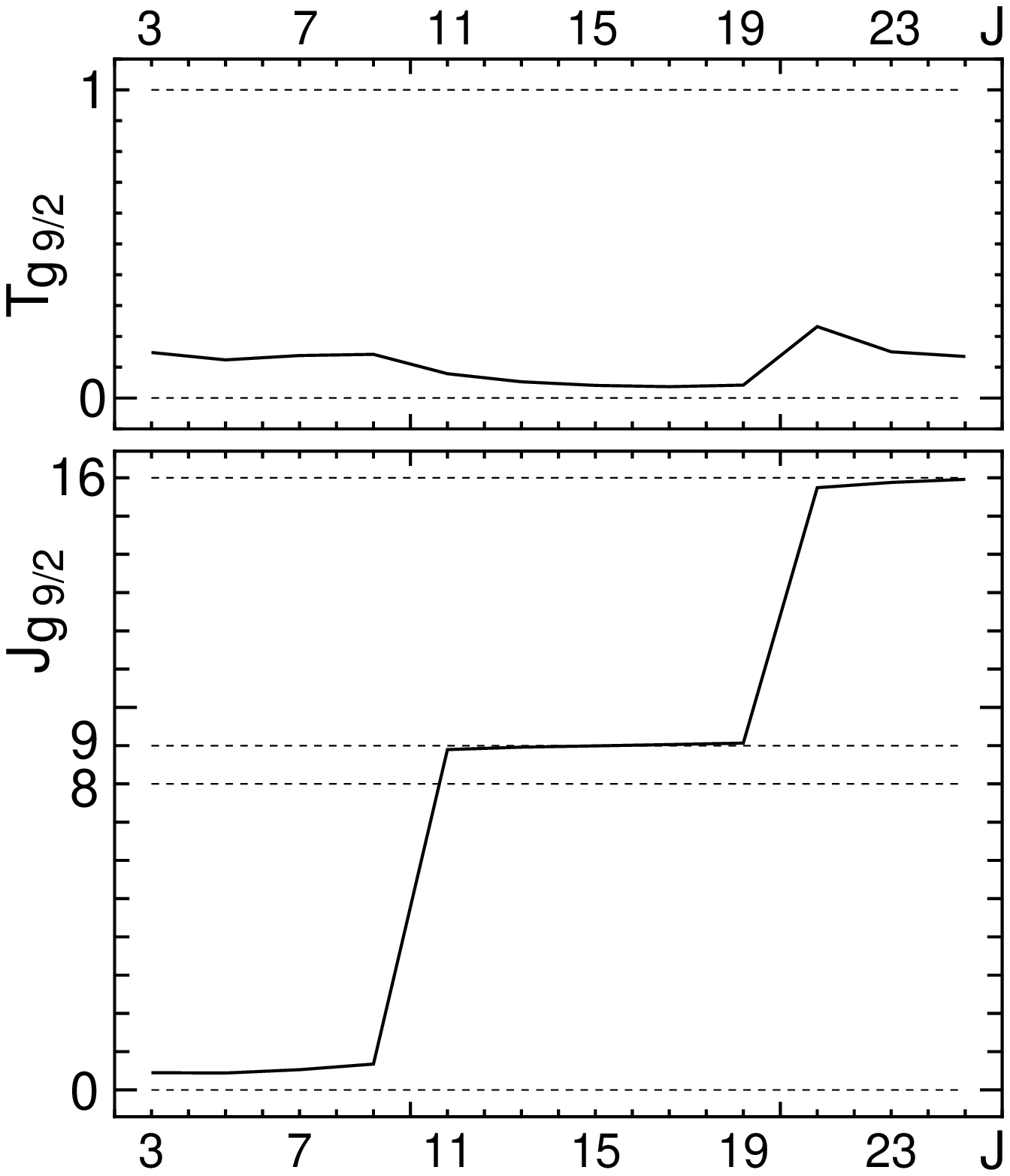}
   \caption{The expectation values of spin and isospin,
            $J_{g9/2}$ and $T_{g9/2}$, for the odd-$J$
            positive-parity yrast states of $^{66}$Ge.}
  \label{fig17}
\end{figure}
%=================================================================

\newpage

%===============  table Ap2  ========================
\begin{table}[h]
\caption{The proton and neutron numbers $\langle n_a^\pi \rangle$ and
         $\langle n_a^\nu \rangle$, and calculated $Q$ moments (in $e$ fm$^2$),
         for the odd-$J$ positive-parity yrast states of $^{68}$Ge.}
\begin{tabular}{c|cccc|cccc|c}   \hline
        & \multicolumn{4}{c}{proton} & \multicolumn{4}{|c|}{neutron} & \\ \hline
$^{68}$Ge & $p_{3/2}$ & $f_{5/2}$ & $p_{1/2}$ & $g_{9/2}$
         & $p_{3/2}$ & $f_{5/2}$ & $p_{1/2}$ & $g_{9/2}$ & $Q$  \\ \hline
  $3_1^+$  & 1.76 & 1.58 & 0.57 & 0.10 & 3.00 & 3.54 & 1.25 & 0.21 & +0.05 \\
  $5_1^+$  & 1.71 & 1.51 & 0.68 & 0.10 & 3.03 & 3.49 & 1.27 & 0.21 & -15.7 \\
  $7_1^+$  & 1.75 & 1.61 & 0.53 & 0.12 & 3.05 & 3.39 & 1.33 & 0.23 &  -8.3 \\ \hline
  $9_1^+$  & 1.61 & 1.60 & 0.64 & 0.16 & 2.22 & 2.93 & 0.85 & 1.99 & -48.9 \\ \hline
 $11_1^+$  & 1.23 & 1.20 & 0.61 & 0.96 & 2.70 & 3.00 & 1.13 & 1.16 & -84.4 \\
 $13_1^+$  & 1.23 & 1.18 & 0.54 & 1.01 & 2.58 & 3.17 & 1.17 & 1.08 & -90.5 \\
 $15_1^+$  & 1.19 & 1.17 & 0.63 & 1.02 & 2.58 & 3.20 & 1.16 & 1.07 & -94.6 \\
 $17_1^+$  & 1.19 & 1.27 & 0.55 & 1.00 & 2.70 & 3.05 & 1.12 & 1.13 & -79.3 \\ \hline
 $19_1^+$  & 1.28 & 1.11 & 0.55 & 1.07 & 1.71 & 2.64 & 0.68 & 2.97 & -86.6 \\
 $21_1^+$  & 1.27 & 1.13 & 0.53 & 1.07 & 1.75 & 2.51 & 0.79 & 2.96 & -89.8 \\
 $23_1^+$  & 1.18 & 1.15 & 0.55 & 1.12 & 1.79 & 2.30 & 1.01 & 2.91 & -92.4 \\
 $25_1^+$  & 1.06 & 1.13 & 0.56 & 1.26 & 1.69 & 2.32 & 1.21 & 2.78 & -93.1 \\ \hline
% $27_1^+$ & 0.18 & 0.87 & 0.13 & 2.82 & 1.76 & 2.26 & 0.78 & 3.20 & -68.9 \\ \hline
\end{tabular}
\label{tableAp2}
\end{table}
%=========================================================

%===============  fig. 18  ========================================
\begin{figure}[h]
\includegraphics[width=6.0cm,height=6.2cm]{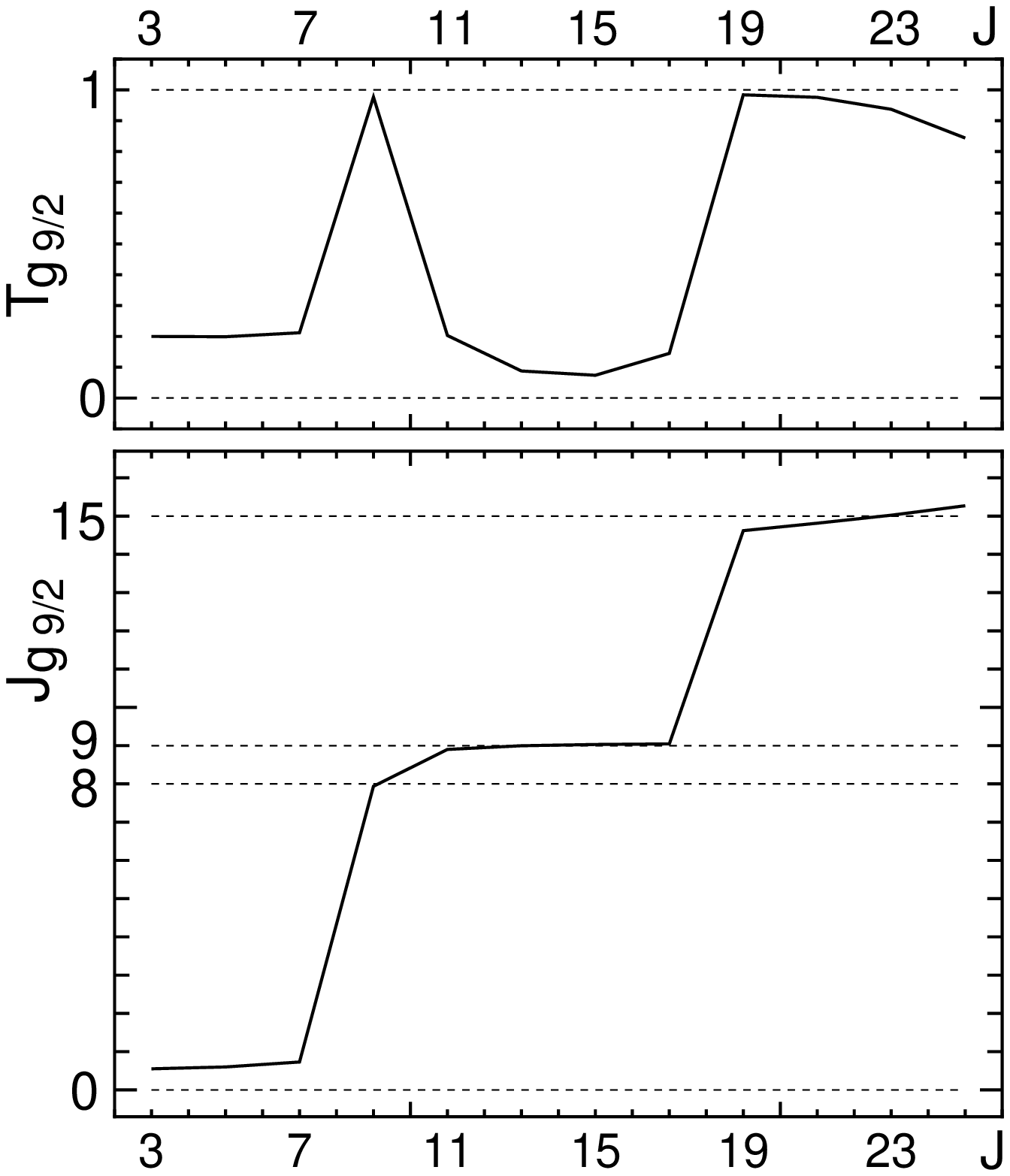}
   \caption{The expectation values of spin and isospin,
            $J_{g9/2}$ and $T_{g9/2}$, for the odd-$J$
            positive-parity yrast states of $^{68}$Ge.}
  \label{fig18}
\end{figure}
%=================================================================

\newpage

\section{Structure of the negative-parity yrast states}

%===============  table Ap3  ========================
\begin{table}[h]
\caption{The proton and neutron numbers $\langle n_a^\pi \rangle$ and
         $\langle n_a^\nu \rangle$, and calculated $Q$ moments (in $e$ fm$^2$),
         for the odd-$J$ negative-parity yrast states and
         some other states of $^{66}$Ge.}
\begin{tabular}{c|cccc|cccc|c}   \hline
        & \multicolumn{4}{c}{proton} & \multicolumn{4}{|c|}{neutron} & \\ \hline
$^{66}$Ge & $p_{3/2}$ & $f_{5/2}$ & $p_{1/2}$ & $g_{9/2}$
         & $p_{3/2}$ & $f_{5/2}$ & $p_{1/2}$ & $g_{9/2}$ & $Q$  \\ \hline
  $3_1^-$  & 1.58 & 1.67 & 0.48 & 0.27 & 2.16 & 2.22 & 0.72 & 0.91 & -37.8 \\
  $5_1^-$  & 1.60 & 1.69 & 0.54 & 0.18 & 2.21 & 2.22 & 0.59 & 0.98 & -48.0 \\
  $7_1^-$  & 1.64 & 1.64 & 0.55 & 0.17 & 1.94 & 2.43 & 0.67 & 0.96 & -47.9 \\
  $9_1^-$  & 1.64 & 1.64 & 0.58 & 0.14 & 1.90 & 2.46 & 0.65 & 0.98 & -51.0 \\
 $11_1^-$  & 1.66 & 1.79 & 0.46 & 0.09 & 1.87 & 2.46 & 0.65 & 1.03 & -46.4 \\
 $13_1^-$  & 1.60 & 1.91 & 0.41 & 0.08 & 1.86 & 2.46 & 0.64 & 1.04 & -49.2 \\ \hline
 $15_1^-$  & 1.29 & 1.14 & 0.55 & 1.02 & 1.50 & 1.89 & 0.60 & 2.01 & -87.7 \\
 $17_1^-$  & 1.29 & 1.15 & 0.55 & 1.02 & 1.55 & 1.75 & 0.69 & 2.01 & -91.1 \\
 $19_1^-$  & 1.29 & 1.15 & 0.53 & 1.03 & 1.36 & 2.04 & 0.61 & 2.00 & -86.7 \\
 $19_2^-$  & 1.22 & 1.19 & 0.58 & 1.01 & 1.56 & 1.45 & 0.98 & 2.01 & -94.6 \\
 $21_1^-$  & 1.28 & 1.14 & 0.55 & 1.03 & 1.32 & 2.08 & 0.61 & 1.99 & -89.0 \\
 $21_2^-$  & 1.20 & 1.18 & 0.59 & 1.03 & 1.44 & 1.39 & 1.17 & 2.00 & -96.5 \\
 $23_1^-$  & 1.26 & 1.14 & 0.58 & 1.03 & 1.31 & 2.09 & 0.58 & 2.01 & -91.1 \\ \hline
 $25_1^-$  & 0.78 & 1.03 & 0.17 & 2.02 & 1.15 & 1.28 & 0.59 & 2.99 & -84.9 \\ \hline
\end{tabular}
\label{tableAp3}
\end{table}
%=========================================================

%===============  fig. 19  ========================================
\begin{figure}[h]
\includegraphics[width=6.0cm,height=7.0cm]{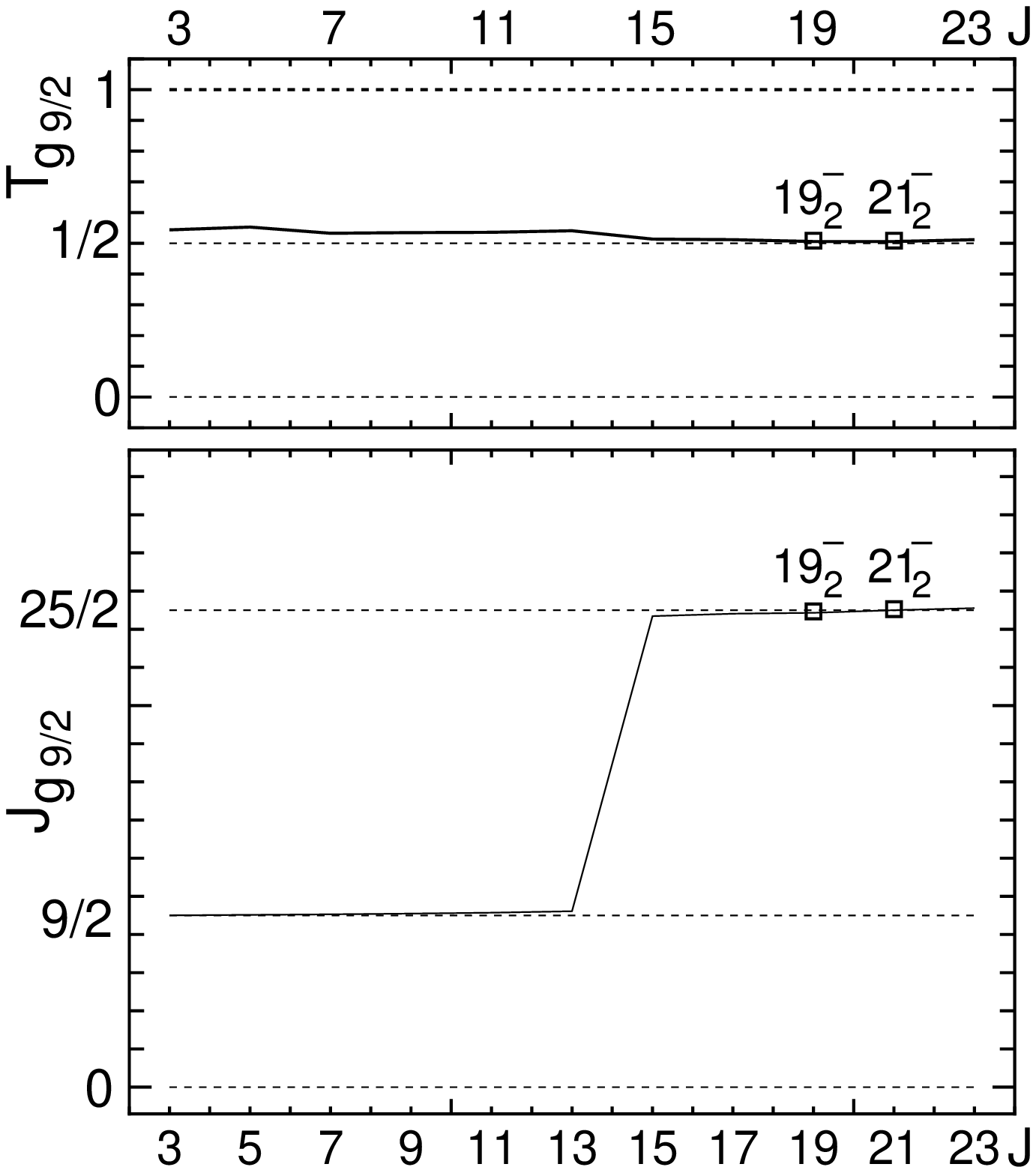}
   \caption{The expectation values of spin and isospin,
            $J_{g9/2}$ and $T_{g9/2}$, for the odd-$J$
            negative-parity yrast states (lines) and some other states (marks)
            of $^{66}$Ge.}
  \label{fig19}
\end{figure}
%=================================================================

\newpage

%===============  table Ap4  ========================
\begin{table}[h]
\caption{The proton and neutron numbers $\langle n_a^\pi \rangle$ and
         $\langle n_a^\nu \rangle$, and calculated $Q$ moments (in $e$ fm$^2$),
         for the odd-$J$ negative-parity yrast states and some other states
         of $^{68}$Ge.}
\begin{tabular}{c|cccc|cccc|c}   \hline
        & \multicolumn{4}{c}{proton} & \multicolumn{4}{|c|}{neutron} & \\ \hline
$^{68}$Ge & $p_{3/2}$ & $f_{5/2}$ & $p_{1/2}$ & $g_{9/2}$
         & $p_{3/2}$ & $f_{5/2}$ & $p_{1/2}$ & $g_{9/2}$ & $Q$  \\ \hline
  $3_1^-$  & 1.61 & 1.61 & 0.51 & 0.28 & 2.57 & 3.35 & 1.12 & 0.96 & -40.5 \\
  $5_1^-$  & 1.60 & 1.65 & 0.55 & 0.20 & 2.57 & 3.37 & 1.04 & 1.03 & -54.8 \\
  $7_1^-$  & 1.67 & 1.60 & 0.61 & 0.13 & 2.69 & 3.20 & 1.04 & 1.06 & -57.7 \\
  $9_1^-$  & 1.67 & 1.55 & 0.68 & 0.10 & 2.76 & 3.04 & 1.10 & 1.11 & -67.5 \\
 $11_1^-$  & 1.57 & 1.39 & 0.94 & 0.10 & 2.78 & 2.94 & 1.18 & 1.12 & -70.9 \\
 $13_1^-$  & 1.45 & 1.60 & 0.87 & 0.08 & 2.91 & 2.78 & 1.18 & 1.13 & -70.2 \\ \hline
 $15_1^-$  & 1.26 & 1.17 & 0.55 & 1.02 & 2.22 & 2.78 & 0.97 & 2.03 & -92.6 \\
 $17_1^-$  & 1.25 & 1.14 & 0.59 & 1.02 & 2.20 & 2.97 & 0.79 & 2.04 & -91.8 \\
 $19_1^-$  & 1.26 & 1.19 & 0.53 & 1.02 & 2.38 & 2.41 & 1.17 & 2.03 & -96.2 \\
 $19_2^-$  & 1.26 & 1.15 & 0.58 & 1.02 & 2.15 & 2.95 & 0.86 & 2.04 & -91.1 \\
 $21_1^-$  & 1.18 & 1.18 & 0.62 & 1.02 & 2.39 & 2.34 & 1.24 & 2.03 & -99.2 \\
 $21_2^-$  & 1.22 & 1.16 & 0.59 & 1.02 & 2.15 & 2.76 & 1.05 & 2.04 & -92.8 \\
 $23_1^-$  & 1.24 & 1.15 & 0.59 & 1.02 & 2.09 & 2.67 & 1.20 & 2.05 & -94.3 \\ \hline
 $25_1^-$  & 0.78 & 0.99 & 0.22 & 2.01 & 1.85 & 2.29 & 0.86 & 3.00 & -85.1 \\ \hline
\end{tabular}
\label{tableAp4}
\end{table}
%=========================================================

%===============  fig. 20  ========================================
\begin{figure}[h]
\includegraphics[width=6.0cm,height=7.0cm]{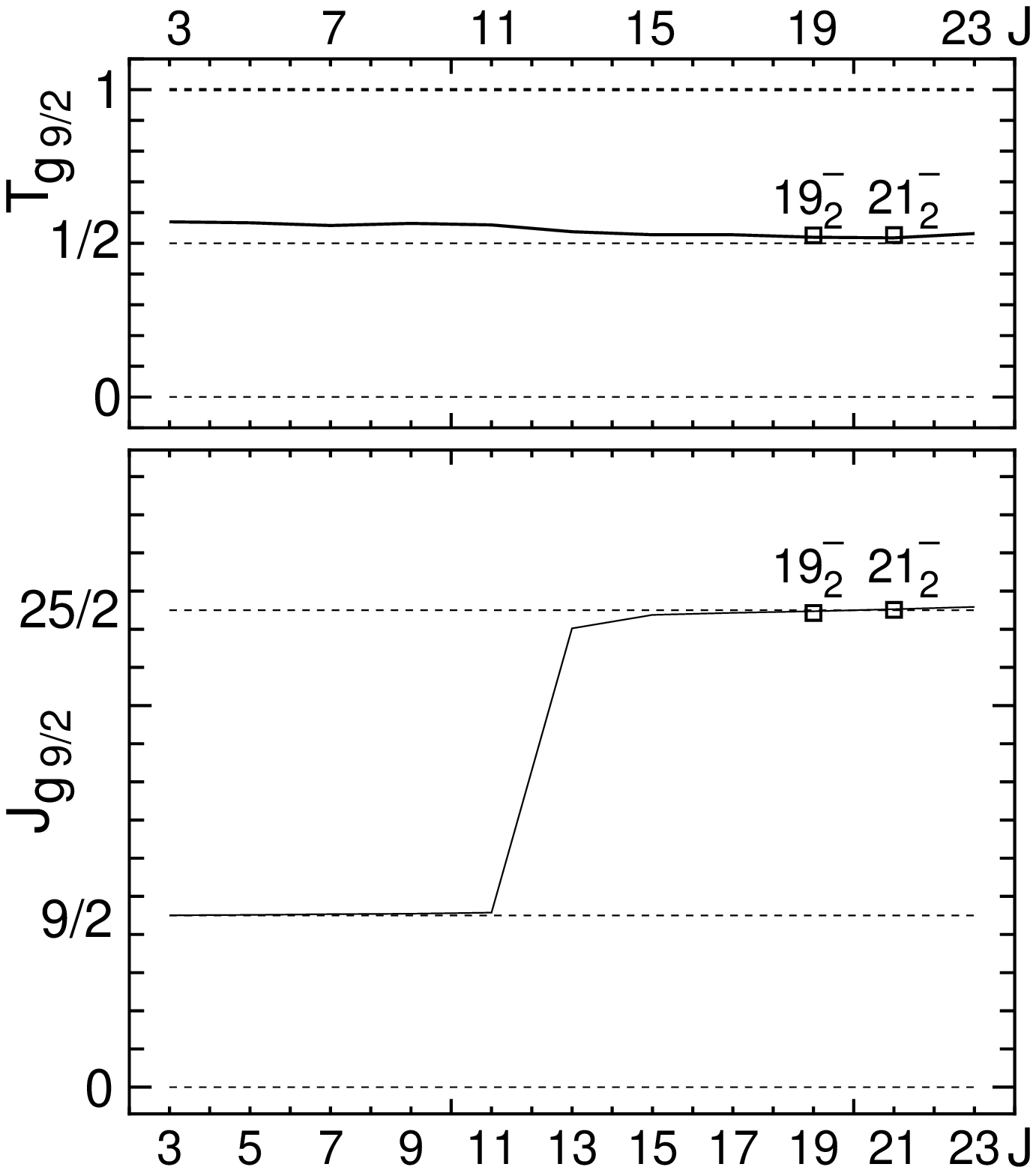}
   \caption{The expectation values of spin and isospin, 
            $J_{g9/2}$ and $T_{g9/2}$, for the odd-$J$
            negative-parity yrast states (lines) and some other states (marks)
            of $^{68}$Ge.}
  \label{fig20}
\end{figure}
%=================================================================

\newpage

%===============  table Ap5  ========================
\begin{table}[t]
\caption{The proton and neutron numbers $\langle n_a^\pi \rangle$ and
         $\langle n_a^\nu \rangle$, and calculated $Q$ moments (in $e$ fm$^2$),
         for the even-$J$ negative-parity yrast states and some other states
         of $^{68}$Ge.}
\begin{tabular}{c|cccc|cccc|c}   \hline
        & \multicolumn{4}{c}{proton} & \multicolumn{4}{|c|}{neutron} & \\ \hline
$^{68}$Ge & $p_{3/2}$ & $f_{5/2}$ & $p_{1/2}$ & $g_{9/2}$
         & $p_{3/2}$ & $f_{5/2}$ & $p_{1/2}$ & $g_{9/2}$ & $Q$  \\ \hline
  $4_1^-$  & 1.70 & 1.75 & 0.47 & 0.08 & 2.79 & 2.87 & 1.24 & 1.11 & -44.0 \\
  $6_1^-$  & 1.70 & 1.73 & 0.49 & 0.08 & 2.65 & 3.12 & 1.15 & 1.09 & -56.7 \\
  $8_1^-$  & 1.69 & 1.52 & 0.69 & 0.10 & 2.65 & 3.15 & 1.10 & 1.09 & -66.0 \\
 $10_1^-$  & 1.51 & 1.27 & 1.13 & 0.09 & 2.54 & 3.20 & 1.18 & 1.08 & -74.5 \\
 $12_1^-$  & 1.63 & 1.61 & 0.67 & 0.09 & 2.76 & 2.97 & 1.14 & 1.13 & -52.7 \\ \hline
 $14_1^-$  & 1.26 & 1.17 & 0.55 & 1.02 & 2.22 & 2.78 & 0.97 & 2.04 & -74.1 \\
 $16_1^-$  & 1.25 & 1.14 & 0.59 & 1.02 & 2.20 & 2.97 & 0.79 & 2.04 & -75.6 \\
 $16_2^-$  & 1.17 & 1.23 & 0.58 & 1.02 & 2.24 & 2.93 & 0.79 & 2.04 & -81.7 \\
 $18_1^-$  & 1.21 & 1.16 & 0.60 & 1.02 & 2.13 & 2.92 & 0.92 & 2.03 & -89.2 \\
 $18_2^-$  & 1.21 & 1.18 & 0.54 & 1.08 & 2.23 & 2.71 & 1.06 & 2.00 & -88.7 \\
 $20_1^-$  & 1.22 & 1.22 & 0.51 & 1.05 & 2.37 & 2.61 & 1.00 & 2.01 & -85.9 \\
 $22_1^-$  & 1.03 & 1.72 & 0.24 & 1.02 & 2.41 & 2.38 & 1.18 & 2.04 & -81.0 \\ \hline
 $24_1^-$  & 0.70 & 0.86 & 0.44 & 2.01 & 1.73 & 2.46 & 0.80 & 3.01 & -86.6 \\
 $26_1^-$  & 0.74 & 1.00 & 0.26 & 2.00 & 1.71 & 2.31 & 0.97 & 3.01 & -88.7 \\
 $28_1^-$  & 0.89 & 1.10 & 0.00 & 2.01 & 1.56 & 2.21 & 1.23 & 3.01 & -90.2 \\ \hline
 $30_1^-$  & 0.03 & 1.00 & 0.00 & 2.97 & 1.42 & 1.99 & 0.55 & 4.03 & -61.8 \\ \hline
\end{tabular}
\label{tableAp5}
\end{table}
%=========================================================

%===============  fig. 21  ========================================
\begin{figure}[h]
\includegraphics[width=6.0cm,height=7.0cm]{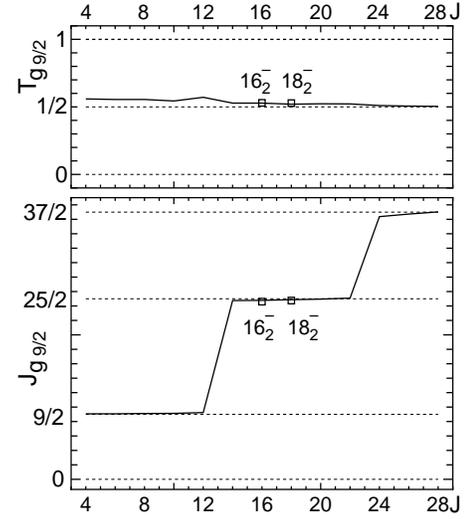}
   \caption{The expectation values of spin and isospin,
            $J_{g9/2}$ and $T_{g9/2}$, for the even-$J$
            negative-parity yrast states (lines) and some other states (marks)
            of $^{68}$Ge.}
  \label{fig21}
\end{figure}
%=================================================================

%\newpage

%------------------------------------------------------------------------------

%------------------------------------------------------------------------------


\begin{thebibliography} {99}

\bibitem{Nolte} E. Nolte, Y. Shida, W. Kutschera, R. Prestele, and H. Morinaga,
 Z. Phys. {\bf 268}, 267 (19742).

\bibitem{Cleenmann} L. Cleenmann, J. Eberth, W. Neumann, W. Wiehl, and V. Zobel,
 Nucl. Phys. {\bf A334}, 157 (1980).

\bibitem{Sound} R. Soundranayagam, R.B. Piercey, A.V. Ramayya, J.H. Hamilton,
 A.Y. Ahmed, H. Yamada, C.F. Maguire, G.L. Bomar, R.L. Robinson, and H.J.Kim,
 Phys. Rev. C {\bf 25}, 1575 (1982).

\bibitem{Boucenna} A. Boucenna, L. Kraus, I. Linck, and Tsan Ung Chan,
 Phys. Rev C {\bf 42}, 1297 (1990).

\bibitem{Hermkens} U. Hermkens, F. Becker, J. Eberth, S. Freund . Mylaeus,
 S. Skoda, W. Teichert, and A.v.d. Werth, Z. Phys. A {\bf 343}, 371 (1992).

\bibitem{Stefanova} E.A. Stefanova {\it et al.},
 Phys. Rev. C {\bf 67}, 054319 (2003).

\bibitem{Pardo} R.C. Pardo, C.N. Davids, M.J. Murphy, E.B. Norman, and L.A. Parks,
  Phys. Rev. C {\bf 15}, 1811 (1977).

\bibitem{Lima} A.P. de Lima {\it et al.}, Phys. Rev. C {\bf 23}, 213 (1981).

\bibitem{Chat1} L. Chaturvedi {\it et al.}, Phys. Rev. C {\bf 43}, 2541 (1991).

\bibitem{Chat2} L. Chaturvedi {\it et al.},
 Int. J. Mod. Phys. E {\bf 5}, 1565 (1996).

\bibitem{Ward} D. Ward {\it et al.}, Phys. Rev. C {\bf 63}, 014301 (2001).

\bibitem{Barclay} M.E. Barclay, J. Phys. G {\bf 12}, L295 (1986).

\bibitem{Petrovici1} A. Petrovici, K.W. Schmid, F. G\"umer, and A. Faessler,
 Nucl. Phys. {\bf A483}, 317 (1988).

\bibitem{Petrovici2} A. Petrovici, K.W. Schmid, F. G\"umer, and A. Faessler,
 Nucl. Phys. {\bf A504}, 277 (1989).

\bibitem{Petrovici3} A. Petrovici, K.W. Schmid, F. G\"umer, and A. Faessler,
 Nucl. Phys. {\bf A517}, 108 (1990).

\bibitem{Hsieh} S.T. Hsieh, H.C. Chiang, and Der-San Chuu,
 Phys. Rev. C {\bf 46}, 195 (1992).

\bibitem{Chuu} Der-San Chuu, S.T. Hsieh, and H.C. Chiang,
 Phys. Rev. C {\bf 47}, 183 (1993).

\bibitem{Elliott} J.P. Elliott, J.A. Evans, V.S. Lac, and G.L. Long,
 Nucl. Phys. {\bf A609}, 1 (1996).

\bibitem{Bengtsson} R. Bengtsson, P. M\"oller, J.R. Nix, and J. Zhang,
 Phys. Scr. {\bf 29}, 402 (1984).

\bibitem{Nazarewicz} W. Nazarewicz, J. Dudek, R. Bengtsson, T. Begtsson,
 and I. Ragnarsson, Nucl. Phys. {\bf A435}, 397 (1985).

\bibitem{Sarri} P. Sarriguren, E. Moya de Guerra, and A. Escuderos,
 Nucl. Phys. {\bf A658}, 13 (1999).

\bibitem{Ennis} P.J. Ennis, C.J. Lister, W. Gelletly, H.G. Price, B.J. Varley,
 P.A. Butler, T. Hoare, S. Cwiok, and W. Nazarewicz,
 Nucl. Phys. {\bf A535}, 392 (1991).

\bibitem{Yamagami} M. Yamagami, K. Matsuyanagi, and M. Matsuo,
 Nucl. Phys. {\bf A693}, 579 (2001).

\bibitem{Kaneko}  K. Kaneko, M. Hasegawa, and T. Mizusaki,
 Phys. Rev. C {\bf 66}, 051306(R) (2002).

\bibitem{Hase1} M. Hasegawa and K. Kaneko, Phys. Rev. C {\bf 59}, 1449 (1999).

\bibitem{Hase2} M. Hasegawa, K. Kaneko, and S. Tazaki,
 Nucl. Phys. A {\bf 674}, 411 (2000); {\bf688}, 765 (2001);
 Prog. Theor. Phys. {\bf 107}, 731 (2002).

\bibitem{Mizusaki} T. Mizusaki, RIKEN Accel. Prog. Rep. {\bf 33}, 14 (2000).

\bibitem{Hase4} M. Hasegawa, K. Kaneko, and T. Mizusaki, Phys. Rev. C {\bf 70},
 031301(R) (2004).

\bibitem{Rudolph} D. Rudolph {\it et al.}, Eur. Phys. J. A {\bf 6}, 377 (1999).

\bibitem{Aberg} A. Juodagalvis and S. ${\rm \AA}$berg, Nucl. Phys. {\bf A683},
 206(2001).

\bibitem{Sun2} Yang Sun, Jing-ye Zhang, M. Guidry, Jie Meng, and Soojae Im,
 Phys. Rev. C {\bf 62}, 021601(R) (2000).

\bibitem{ENSDF} http://www.nndc.bnl.gov/nndc/ensdf

\bibitem{Hase3} M. Hasegawa, K. Kaneko, T. Mizusaki, and S. Tazaki,
 Phys. Rev. C {\bf 69}, 034324 (2004).

\bibitem{Kaneko2}  K. Kaneko, M. Hasegawa, and T. Mizusaki,
 submitted to Phys. Rev. C.

\bibitem{Mizusaki2}  T. Mizusaki, T. Otsuka, Y. Utsuno, M. Honma, and T. Sebe,
 Phys. Rev. C {\bf 59}, R1846 (1999).

\bibitem{Hara}  K. Hara, Yang Sun, and T. Mizusaki,
 Phys. Rev. Lett. {\bf 83}, 1922 (1999).


\end{thebibliography}
\end{document}